\documentclass[12pt]{article}
\usepackage{epsfig}

\usepackage{amsmath}
\usepackage{feynarts,axodraw,graphicx}
\numberwithin{equation}{section}

\newcommand{\re}{\mathop{\mathrm{Re}}\nolimits}
\newcommand{\li}{\mathop{\mathrm{Li}_2}\nolimits}

\catcode`@=11
\newcount\@tempcntc
\def\@citex[#1]#2{\if@filesw\immediate\write\@auxout{\string\citation{#2}}\fi
  \@tempcnta\z@\@tempcntb\m@ne\def\@citea{}\@cite{\@for\@citeb:=#2\do
    {\@ifundefined
       {b@\@citeb}{\@citeo\@tempcntb\m@ne\@citea\def\@citea{,}{\bf ?}\@warning
       {Citation `\@citeb' on page \thepage \space undefined}}%
    {\setbox\z@\hbox{\global\@tempcntc0\csname b@\@citeb\endcsname\relax}%
     \ifnum\@tempcntc=\z@ \@citeo\@tempcntb\m@ne
       \@citea\def\@citea{,}\hbox{\csname b@\@citeb\endcsname}%
     \else
      \advance\@tempcntb\@ne
      \ifnum\@tempcntb=\@tempcntc
      \else\advance\@tempcntb\m@ne\@citeo
      \@tempcnta\@tempcntc\@tempcntb\@tempcntc\fi\fi}}\@citeo}{#1}}
\def\@citeo{\ifnum\@tempcnta>\@tempcntb\else\@citea\def\@citea{,}%
  \ifnum\@tempcnta=\@tempcntb\the\@tempcnta\else
   {\advance\@tempcnta\@ne\ifnum\@tempcnta=\@tempcntb \else \def\@citea{--}\fi
    \advance\@tempcnta\m@ne\the\@tempcnta\@citea\the\@tempcntb}\fi\fi}
\catcode`@=12

\begin{document}
\thispagestyle{empty}
\title{
\vskip-3cm{\baselineskip14pt
\centerline{\normalsize DESY 07-103\hfill ISSN 0418-9833}
\centerline{\normalsize MPP-2007-96\hfill}
\centerline{\normalsize July 2007\hfill}}
\vskip1.5cm
\boldmath
\bf Electroweak corrections to $W$-boson hadroproduction at finite transverse
momentum
\unboldmath}

\author{W. Hollik${}^{\rm a}$, T. Kasprzik${}^{\rm a}$,
B.A. Kniehl${}^{\rm b}$\bigskip
\\
{\normalsize\it ${}^{\rm a}$ Max-Planck-Institut f\"ur Physik
(Werner-Heisenberg-Institut),}\\
{\normalsize\it F\"ohringer Ring 6, 80805 Munich, Germany}\bigskip
\\
{\normalsize\it ${}^{\rm b}$ II. Institut f\"ur Theoretische Physik,
Universit\"at Hamburg,}\\
{\normalsize\it Luruper Chaussee 149, 22761 Hamburg, Germany}}

\date{}

\maketitle

\begin{abstract}
We calculate the full one-loop electroweak radiative corrections to the
cross section of single $W$-boson inclusive hadroproduction at finite
transverse momentum ($p_T$).
This includes the $\mathcal{O}(\alpha)$ corrections to $W+j$ production, the
$\mathcal{O}(\alpha_s)$ corrections to $W+\gamma$ production, and the
tree-level contribution from $W+j$ photoproduction with one direct or resolved
photon in the initial state.
We present the integrated cross section as a function of a minimum-$p_T$
cut as well as the $p_T$ distribution for the experimental conditions at
the Fermilab Tevatron and the CERN LHC and estimate the theoretical
uncertainties.
\medskip

\end{abstract}

\newpage

\section{Introduction}

The hadroproduction of single $W$ bosons via the Drell-Yan process in
$p\overline{p}$ collisions at the CERN S$p\overline{p}$S led to the discovery
of this particle in 1983 \cite{Arnison:1983rp}.
Nowadays, this process serves as a standard candle to calibrate and monitor
the luminosity of hadronic collisions, since its cross section is rather
sizeable and $W$ bosons are straightforward to identify experimentally thanks
to their simple and distinct decay signature. 
The quality of the luminosity determination is thus limited by the precision
to which this cross section is predicted theoretically.
It is, therefore, mandatory to calculate higher-order radiative corrections.
At present, they are known at next-to-leading order (NLO)
\cite{Kubar-Andre:1978uy} and next-to-next-to-leading order (NNLO)
\cite{Hamberg:1990np} in quantum chromodynamics (QCD) as well as at NLO
\cite{Baur:1998kt} in the electroweak sector of the standard model (SM) of
elementary particle physics.

In order for the $W$ boson to acquire finite transverse momentum ($p_T$), it
must be produced in association with one or more other particles.
To lowest order (LO) in QCD, the additional particle is a gluon ($g$), quark
($q$), or antiquark ($\overline{q}$), materialising as a hadron jet ($j$).
The corresponding partonic subprocesses are of $\mathcal{O}(\alpha\alpha_s)$.
Their cross sections are presently known at NLO \cite{Ellis:1981hk} and NNLO
\cite{Anastasiou:2003ds} in QCD, i.e.\ at $\mathcal{O}(\alpha\alpha_s^2)$ and
$\mathcal{O}(\alpha\alpha_s^3)$, respectively.
Very recently, also the one-loop electroweak corrections, at
$\mathcal{O}(\alpha^2\alpha_s)$, were considered \cite{Kuhn:2007qc}.
This is also the topic of the present paper.
However, as explained below, we actually study somewhat different cross
section observables and arrange for our results to be manifestly infrared (IR)
safe by avoiding kinematic cuts that destroy the inclusiveness of massless
quanta.
In fact, the situation is complicated by the circumstance that both photons
and gluons can appear as bremsstrahlung.

The observable we thus wish to investigate is the differential cross section
for the inclusive hadroproduction of single $W$ bosons with finite $p_T$ at
$\mathcal{O}(\alpha^2\alpha_s)$.
Specifically, we concentrate on the $p_T$ distribution and the integrated
cross section as a function of a minimum-$p_T$ cut, leaving the distributions
in other observables, such as rapidity, for future work.
At LO, the system $X$ recoiling against the $W$ boson is purely hadronic,
while at $\mathcal{O}(\alpha^2\alpha_s)$ it can include a photon ($\gamma$).
We thus also have to consider $W+\gamma$ production, whose LO partonic cross
sections are of $\mathcal{O}(\alpha^2)$, because its NLO QCD correction
contributes at the very order we are aiming at.
In fact, the real radiative corrections to $W+j$ and $W+\gamma$ production
receive contributions from a common set of $2\to3$ partonic subprocesses.
In the former (latter) case, IR singularities from the radiation of soft
photons (gluons) cancel against similar contributions from virtual photons
(gluons) by the Bloch-Nordsieck theorem \cite{BloNo}.
Similarly, the IR singularities from collinear final-state radiation (FSR)
cancel against similar contributions from the virtual corrections by the
Kinoshita-Lee-Nauenberg theorem \cite{Kinoshita:1962ur}.
The residual IR singularities from collinear initial-state radiation (ISR) are
factorised and absorbed at $\mathcal{O}(\alpha)$ ($\mathcal{O}(\alpha_s)$)
into the parton distribution functions (PDFs).
This procedure leads to manifestly IR-safe cross section observables and
translates into unique and simple event selection criteria on the experimental
side.

By contrast, the notion of electroweak NLO corrections to $W+j$ production
comprises a conceptual problem.
In fact, in the treatment of final-state collinear singularities caused by the
parallel emission of a photon from an outgoing (anti)quark line, one is led to
introduce a cut in an appropriate separation variable.
Within the collinear phase space region thus defined, the (anti)quark-photon
system is effectively treated as one particle whose momentum is identified
with that of $j$ and thus subject to an acceptance cut in transverse momentum,
$p_T(j)>p_T^\mathrm{min}(j)$, to ensure the experimental observation of $j$.
This includes phase space configurations where the photon essentially carries
all the momentum, while the (anti)quark can, in principle, be arbitrarily
soft.
This will not generate any soft IR singularities.
However, since (anti)quark and gluon jets can, in general, not be
distinguished experimentally on an event-by-event basis, the same
recombination procedure needs to be applied to a gluon-photon system in the
final state as well.
This time, a soft gluon will inevitably produce an IR singularity, which can
only be canceled by the NLO QCD corrections to $W+\gamma$ production, so that
one falls back to the symmetric procedure outlined in the preceding
paragraph.
Formally, this soft-gluon singularity can be avoided by applying the 
$p_T^\mathrm{min}(j)$ cut just to the transverse momentum of the gluon, even
if it is accompanied by a collinear photon.
However, such a prescription is purely academic and quite unsuitable for
experimental implementation because (anti)quark and gluon jets are treated on
different footings.

By crossing external lines, the LO partonic subprocesses of $W+\gamma$
hadroproduction can be converted to those of $W+j$ photoproduction with one
incoming photon participating directly in the hard scattering (direct
photoproduction).
The emission of photons off the proton can happen either elastically or
inelastically, i.e.\ the proton stays intact or is destroyed, respectively.
In both cases, an appropriate PDF can be evaluated in the
Weizs\"acker-Williams approximation \cite{kni,gsv,Martin:2004dh}.
Since they are of $\mathcal{O}(\alpha)$, these direct photoproduction
contributions are of $\mathcal{O}(\alpha^3)$.
Incoming photons can participate in the hard scattering also through their
quark and gluon content, leading to resolved photoproduction.
The contributions from direct and resolved photoproduction are formally of the
same order in the perturbative expansion.
This may be understood by observing that the PDFs of the photon have a leading
behaviour proportional to
$\alpha\ln(M^2/\Lambda_\mathrm{QCD}^2)\propto\alpha/\alpha_s$, where $M$ is
the factorisation scale and $\Lambda_\mathrm{QCD}$ is the asymptotic scale
parameter of QCD.
Although photoproduction contributions are parametrically suppressed by a
factor of $\alpha/\alpha_s$ relative to the $\mathcal{O}(\alpha^2\alpha_s)$
corrections discussed above, we shall include them in our analysis because
they turn out to be quite sizeable in an extensive region of phase space.

This paper is organised as follows.
In Section~\ref{sec:two}, we list the partonic cross sections at LO and
explain how to evaluate the hadronic cross section from them.
In Section~\ref{sec:three}, we discuss in detail the structure of the NLO
corrections.
In Section~\ref{sec:four}, we present our numerical results for
$p\overline{p}$ collisions with centre-of-mass (c.m.) energy
$\sqrt{S}=1.96$~TeV at the Fermilab Tevatron and $pp$ collisions with
$\sqrt{S}=14$~TeV at the CERN Large Hadron Collider (LHC).

\section{Conventions and LO results}
\label{sec:two}

We consider the hadronic process
\begin{equation}
A(p_A)+B(p_B)\to W(p)+X,
\label{eq:had}
\end{equation}
where the four-momentum assignments are indicated in parentheses.
We work in the collinear parton model of QCD \cite{Partonmodell} with $n_f=5$
massless quark flavours $q=u,d,s,c,b$, neglect the masses of the incoming
hadrons, $A$ and $B$, and impose the acceptance cut $p_T>p_T^\mathrm{cut}$ on
the transverse momentum $p_T$ of the $W$ boson.  
(We assign masses to the partons $\gamma,g,q,\overline{q}$ only to regulate
soft and collinear IR singularities in intermediate steps of our calculation.)

Specifically, denoting $u_1=u$, $u_2=c$, $d_1=d$, $d_2=s$, and $d_3=b$,
the relevant partonic subprocesses include
\begin{eqnarray}
u_i+\overline{d}_j&\to&W^++g,
\label{eq:udg}\\
u_i+g&\to&W^++d_j,
\label{eq:ugd}\\
\overline{d}_j+g&\to&W^++\overline{u}_i,
\label{eq:dgu}
\end{eqnarray}
at $\mathcal{O}(\alpha\alpha_s)$,
\begin{eqnarray}
u_i+\overline{d}_j&\to&W^++\gamma,
\label{eq:udp}\\
u_i+\gamma&\to&W^++d_j,
\label{eq:upd}\\
\overline{d}_j+\gamma&\to&W^++\overline{u}_i,
\label{eq:dpu}
\end{eqnarray}
at $\mathcal{O}(\alpha^2)$, and
\begin{eqnarray}
u_i+\overline{d}_j&\to&W^++g+\gamma,
\label{eq:udgp}\\
u_i+g&\to&W^++d_j+\gamma,
\label{eq:ugdp}\\
\overline{d}_j+g&\to&W^++\overline{u}_i+\gamma,
\label{eq:dgup}
\end{eqnarray}
at $\mathcal{O}(\alpha^2\alpha_s)$.
The partonic subprocesses involving a $W^-$ boson emerge through charge
conjugation.
Processes~(\ref{eq:udg})--(\ref{eq:udp}) must be treated also at one loop,
$\mathcal{O}(\alpha^2\alpha_s)$.
Processes~(\ref{eq:upd}) and (\ref{eq:dpu}) contribute to direct
photoproduction and processes~(\ref{eq:udg})--(\ref{eq:dgu}) to resolved
photoproduction.
Since photon emission off protons happens at $\mathcal{O}(\alpha)$, it is
sufficient to deal with photoproduction at tree level.
In summary, we calculate the cross section of process~(\ref{eq:had}) at NLO as
the sum
\begin{equation}
\sigma^{AB\to WX}=\sigma_0^{Wj}+\sigma_0^{W\gamma}
+\sigma_{\mathcal{O}(\alpha)}^{Wj}+\sigma_{\mathcal{O}(\alpha_s)}^{W\gamma}
+\sigma_0^{Wj\gamma}+\sigma_\gamma^{Wj},
\label{eq:sum}
\end{equation}
where $\sigma_0^{Wj}$ and $\sigma_{\mathcal{O}(\alpha)}^{Wj}$ are due to
processes~(\ref{eq:udg})--(\ref{eq:dgu}) at tree level and one loop,
$\sigma_0^{W\gamma}$ and $\sigma_{\mathcal{O}(\alpha_s)}^{W\gamma}$ are due to
process~(\ref{eq:udp}) at tree level and one loop,
$\sigma_0^{Wj\gamma}$ is due to processes~(\ref{eq:udgp})--(\ref{eq:dgup}) at
tree level, and 
$\sigma_\gamma^{Wj}$ is due to processes~(\ref{eq:upd}) and (\ref{eq:dpu}) via
direct photoproduction and due to processes~(\ref{eq:udg})--(\ref{eq:dgu}) via
resolved photoproduction, both at tree-level.

The minimum-$p_T$ cut is necessary to stay away from the regions of phase
space that are sensitive to the collinear IR singularities due to the
$q\to\gamma/g+q^\ast$, $\overline{q}\to\gamma/g+\overline{q}^\ast$,
$\gamma/g\to q+\overline{q}^\ast$, and $\gamma/g\to\overline{q}+q^\ast$
splittings, which are present already at LO.
Here, an asterisk marks a virtual parton.
The cross section $\sigma^{AB\to WX}$ of the hadronic process~(\ref{eq:had})
is related to the cross sections $\hat\sigma^{ab\to Wc(d)}$ of the partonic
subprocesses,
\begin{equation}
a(p_a)+b(p_b)\to W(p)+c(p_c)(+d(p_d)),
\label{eq:subproc}
\end{equation}
where $a,b,c,d=\gamma,g,q,\overline{q}$ and $p_a=x_ap_A$, $p_b=x_bp_B$ with
scaling parameters $x_a$, $x_b$, as the incoherent sum
\begin{equation}
\sigma^{AB\to WX}(S,p_T>p_T^\mathrm{cut})=\sum_{a,b,c(,d)}
\int_{\tau_0}^1\mathrm{d}\tau\,\mathcal{L}_{ab}^{AB}(\tau)
\hat\sigma^{ab\to Wc(d)}(s,p_T>p_T^\mathrm{cut}),
\label{sigmahad}
\end{equation}
where $S=(p_A+p_B)^2$ and $s=(p_a+p_b)^2=\tau S$ are the hadronic and
partonic c.m.\ energies, respectively, $\tau=x_ax_b$, and
\begin{equation}
\mathcal{L}_{ab}^{AB}(\tau)=\int_{\tau}^1\frac{\mathrm{d}x_a}{x_a}
f_{a/A}(x_a,M^2)f_{b/B}\left(\frac{\tau}{x_a},M^2\right)
\end{equation}
is the parton luminosity defined in terms of the PDFs $f_{a/A}(x_a,M^2)$,
$f_{b/B}(x_b,M^2)$.
Here, $M$ denotes the factorisation mass scale.
Introducing the short-hand notation $w=M_W^2$, we have
\begin{equation}
\tau_0=\frac{\left(p_T^\mathrm{cut}+\sqrt{w+(p_T^\mathrm{cut})^2}\right)^2}
{S}.
\label{eq:tau}
\end{equation}

In order to obtain $\hat\sigma^{ab\to Wc(d)}$, we have to evaluate the
transition matrix elements $\mathcal{T}^{ab\to Wc(d)}$ of
processes~(\ref{eq:subproc}), square them, average them over the initial-state
spins and colours, and sum them over the final-state ones, which leads to
$\overline{|\mathcal{T}^{ab\to Wc(d)}|^2}$.
To the order of our calculation, $\mathcal{T}^{ab\to Wcd}$ is calculated at
tree level, while $\mathcal{T}^{ab\to Wc}$ may receive also one-loop
contributions, 
$\mathcal{T}^{ab\to Wc}=\mathcal{T}_0^{ab\to Wc}+\mathcal{T}_1^{ab\to Wc}$,
so that
\begin{equation}
\overline{\left|\mathcal{T}^{ab\to Wc}\right|^2}
=\overline{\left|\mathcal{T}_0^{ab\to Wc}\right|^2}
+2\re\overline{\left[\left(\mathcal{T}_0^{ab\to Wc}\right)^*
\mathcal{T}_1^{ab\to Wc}\right]}.
\end{equation}
Then we have to integrate over the partonic phase spaces imposing the
minimum-$p_T$ cut.
In the following two subsections, we describe how this can be conveniently done
for the two- and three-particle final states, respectively.

Since we are dealing with charged-current interactions of quarks, $u_i$ and
$d_j$, the Cabibbo-Kobayashi-Maskawa quark mixing matrix $V_{ij}$ appears.
At tree level, the cross sections of processes~(\ref{eq:udg})--(\ref{eq:dgup})
contain the overall factor $|V_{ij}|^2$, and a part of the one-loop
corrections is proportional to
$V_{ij}^*V_{ij^\prime}V_{i^\prime j^\prime}^*V_{i^\prime j}$,
where $u_{i^\prime}$ and $d_{j^\prime}$ are virtual quarks.
Since we neglect all down-quark masses, we can sum over the indices of the
virtual and outgoing down quarks to trigger the unitarity relation
$\sum_{j=1}^3V_{ij}V_{i^\prime j}^*=\delta_{ii^\prime}$.
In the case of incoming down quarks, we can absorb the residual appearances
of $|V_{ij}|^2$ into a redefinition of their PDFs, as \cite{Kuhn:2007qc}
\begin{equation}
\tilde f_{d_i/A}(x,M^2)=\sum_{j=1}^3\left|V_{ij}\right|^2f_{d_j/A}(x,M^2),
\end{equation}
and similarly for down antiquarks.
Therefore, it is sufficient to calculate the partonic cross sections for the
flavour-diagonal case, with $V_{ij}=\delta_{ij}$.

\subsection{Two-particle final state}

If parton $d$ is absent in process~(\ref{eq:subproc}), we supplement $s$ by
two more Mandelstam variables, $t=(p_a-p)^2$ and $u=(p_b-p)^2$.
Four-momentum conservation implies that $s+t+u=w$, and we have $p_T^2=tu/s$.
The partonic cross section entering Eq.~(\ref{sigmahad}) is evaluated as
\begin{equation}
\hat\sigma^{ab\to Wc}(s,p_T>p_T^\mathrm{cut})=
\int_{p_T^\mathrm{cut}}^{p_T^\mathrm{max}}\mathrm{d}p_T
\frac{\mathrm{d}\hat\sigma^{ab\to Wc}}{\mathrm{d}p_T},
\end{equation}
where $p_T^\mathrm{max}=(s-w)/(2\sqrt{s})$ and 
\begin{equation}
\frac{\mathrm{d}\hat\sigma^{ab\to Wc}}{\mathrm{d}p_T}=
\frac{p_T}{8\pi s\sqrt{(s-w)^2-4sp_T^2}}
\overline{\left|\mathcal{T}^{ab\to Wc}\right|^2}+(t\leftrightarrow u).
\end{equation}

For the reader's convenience, we list the differential cross sections of
processes~(\ref{eq:udg})--(\ref{eq:dpu}), in the conventional form
\begin{equation}
\frac{\mathrm{d}\hat\sigma^{ab\to Wc}}{\mathrm{d}t}=
\frac{1}{16\pi s^2}\overline{\left|\mathcal{T}^{ab\to Wc}\right|^2},
\end{equation}
at LO.
The Feynman diagrams contributing to processes~(\ref{eq:udg}) and
(\ref{eq:udp}) are displayed in Figs.~\ref{DiagBornudWg} (a) and (b),
respectively.
We have
\begin{eqnarray}
\frac{\mathrm{d}\hat\sigma^{u\overline{d}\to W^+g}}{\mathrm{d}t}&=&
\frac{2\pi\alpha\alpha_s}{9s_w^2}\,\frac{s^2+w^2-2tu}{s^2tu},
\nonumber\\
\frac{\mathrm{d}\hat\sigma^{u\overline{d}\to W^+\gamma}}{\mathrm{d}t}&=&
\frac{\alpha}{12\alpha_s}\left(1+\frac{3t}{s-w}\right)^2
\frac{\mathrm{d}\hat\sigma^{u\overline{d}\to W^+g}}{\mathrm{d}t},
\label{eq:loxs}
\end{eqnarray}
where $s_w=\sin\theta_w$ is the sine of the weak-mixing angle.
Since the $W$-boson mass sets the renormalisation scale of the couplings, it
is natural to adopt the definition of Sommerfeld's fine-structure constant
$\alpha$ in terms of Fermi's constant $G_F$,
\begin{equation}
\alpha=\frac{\sqrt2}{\pi}G_Fs_w^2w.
\end{equation}
The implementation of this renormalisation scheme at one loop is explained in
Section~\ref{sec:udg}.
The cross sections of processes~(\ref{eq:ugd}), (\ref{eq:dgu}),
(\ref{eq:upd}), and (\ref{eq:dpu}) may be obtained from Eq.~(\ref{eq:loxs}) by
exploiting crossing symmetries, as
\begin{eqnarray}
s^2\frac{\mathrm{d}\hat\sigma^{ug\to W^+d}}{\mathrm{d}t}&=&
-\frac{3}{8}\left[s^2\frac{\mathrm{d}\hat\sigma^{u\overline{d}\to W^+g}}
{\mathrm{d}t}\right]_{s\leftrightarrow u},
\nonumber\\
s^2\frac{\mathrm{d}\hat\sigma^{\overline{d}g\to W^+\overline{u}}}{\mathrm{d}t}
&=&
\left[s^2\frac{\mathrm{d}\hat\sigma^{ug\to W^+d}}{\mathrm{d}t}
\right]_{s\leftrightarrow t},
\nonumber\\
s^2\frac{\mathrm{d}\hat\sigma^{u\gamma\to W^+d}}{\mathrm{d}t}&=&
-3\left[s^2\frac{\mathrm{d}\hat\sigma^{u\overline{d}\to W^+\gamma}}
{\mathrm{d}t}\right]_{s\leftrightarrow u},
\nonumber\\
s^2\frac{\mathrm{d}\hat\sigma^{\overline{d}\gamma\to W^+\overline{u}}}
{\mathrm{d}t}&=&
\left[s^2\frac{\mathrm{d}\hat\sigma^{u\gamma\to W^+d}}{\mathrm{d}t}
\right]_{s\leftrightarrow t}.
\end{eqnarray}

\subsubsection{Three-particle final states}
\label{sec:tpfs}

If parton $d$ is present in process~(\ref{eq:subproc}), then the partonic
cross section entering Eq.~(\ref{sigmahad}) may be obtained through a
four-fold phase-space integration along the lines of Ref.~\cite{DittmDoktor}.
We work in the partonic c.m.\ frame and choose our coordinate system so that
$\vec{p_a}$ points along the $z$ direction and $\vec{p_d}$ lies in the $x$-$y$
plane.
We denote the polar angle of $\vec{p_d}$ by $\vartheta$ and the azimuthal
angle of $\vec{p_c}$ by $\varphi$.
As the first three independent variables, we select $p_d^0$, $\vartheta$, and
$\varphi$, which take the values
\begin{equation}
0<p_d^0<\frac{s-w}{2\sqrt{s}},\qquad
0<\vartheta<\pi,\qquad
0<\varphi<2\pi.
\end{equation}
In the case of process~(\ref{eq:udgp}), which contains two massless gauge
bosons in the final state, it is convenient to take the fourth
variable to be $p_c^0$, with values
\begin{equation}
\frac{1}{2}\left(\sqrt{s}-2p_d^0-\frac{w}{\sqrt{s}}\right)<p_c^0<
\frac{1}{2}\left(\sqrt{s}-\frac{w}{\sqrt{s}-2p_d^0}\right).
\end{equation}
We then have
\begin{equation}
\left.\frac{\mathrm{d}^4\hat{\sigma}^{u\overline{d}\to W^+g\gamma}}
{\mathrm{d}p_c^0\mathrm{d}p_d^0\mathrm{d}\cos\vartheta\mathrm{d}\varphi}
\right|_{p_T>p_T^\mathrm{cut}}=
\frac{1}{8(2\pi)^4}\overline{\left|\mathcal{T}^{u\overline{d}\to W^+g\gamma}
\right|^2}\theta\left(p_T-p_T^\mathrm{cut}\right).
\end{equation}
On the other hand, in the case of processes~(\ref{eq:ugdp}) and
(\ref{eq:dgup}), which only contain one massless gauge boson in the final
state, it is more useful to choose the fourth variable to be the angle $\psi$
enclosed between $\vec{p}_c$ and $\vec{p}_d$, with values
\begin{equation}
0<\psi<\pi.
\end{equation}
We then have
\begin{equation}
\left.\frac{\mathrm{d}^4\hat{\sigma}^{ug\to W^+d\gamma}}
{\mathrm{d}p_d^0\mathrm{d}\cos\vartheta\mathrm{d}\varphi\mathrm{d}\psi}
\right|_{p_T>p_T^\mathrm{cut}}=
\frac{p_d^0\left[\sqrt{s}\left(\sqrt{s}-2p_d^0\right)-w\right]}
{16(2\pi)^4\left[\sqrt{s}-2p_d^0\sin^2(\psi/2)\right]^2}
\overline{\left|\mathcal{T}^{ug\to W^+d\gamma}
\right|^2}\theta\left(p_T-p_T^\mathrm{cut}\right),
\end{equation}
and similarly for process~(\ref{eq:dgup}).
In order to implement the minimum-$p_T$ cut, $p_T$ needs to be expressed in
terms of the integration variables, which is conveniently done with the help
of Eqs.~(5.40) and (5.42) of Ref.~\cite{DittmDoktor} and starting from
\begin{equation}
p_T=\sqrt{\left(p_c^1+p_d^1\right)^2+\left(p_c^2+p_d^2\right)^2}.
\end{equation}

\section{NLO results}
\label{sec:three}

We now describe the calculation of the NLO contributions
$\sigma_{\mathcal{O}(\alpha)}^{Wj}$,
$\sigma_{\mathcal{O}(\alpha_s)}^{W\gamma}$, and $\sigma_0^{Wj\gamma}$ of
Eq.~(\ref{eq:sum}) in some detail.

We employ the following tools.
We generate the relevant Feynman diagrams using the symbolic program package
FeynArts \cite{FeynArts}, carry out the spin and colour sums using the program
package FormCalc \cite{FormCalc}, and perform the Passarino-Veltman reduction
of the tensor one-loop integrals \cite{PaVe} using the program package
FeynCalc \cite{FeynCalc}.
Subsequently, we implement the analytical results in a Fortran program.
We evaluate the standard scalar one-loop integrals contained in the purely
weak corrections using the program package LoopTools \cite{LoopTools}, which
incorporates the program library FF \cite{FF}.
The numerical integrations are performed using the program package Cuba
\cite{Cuba}, which provides several different integration routines and is,
therefore, also well suited for cross checks.

\subsection{Virtual electroweak corrections to $W+j$ production}
\label{sec:udg}

\begin{figure}[ht]
\begin{center}
\input{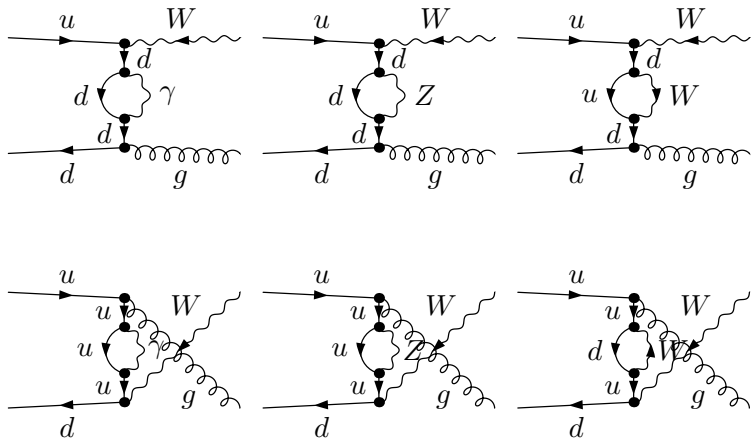}
\caption{$\mathcal{O}(\alpha)$ self-energy diagrams of process~(\ref{eq:udg}).}
\label{selfenudWg}
\end{center}
\end{figure}

\begin{figure}[ht]
\begin{center}
\input{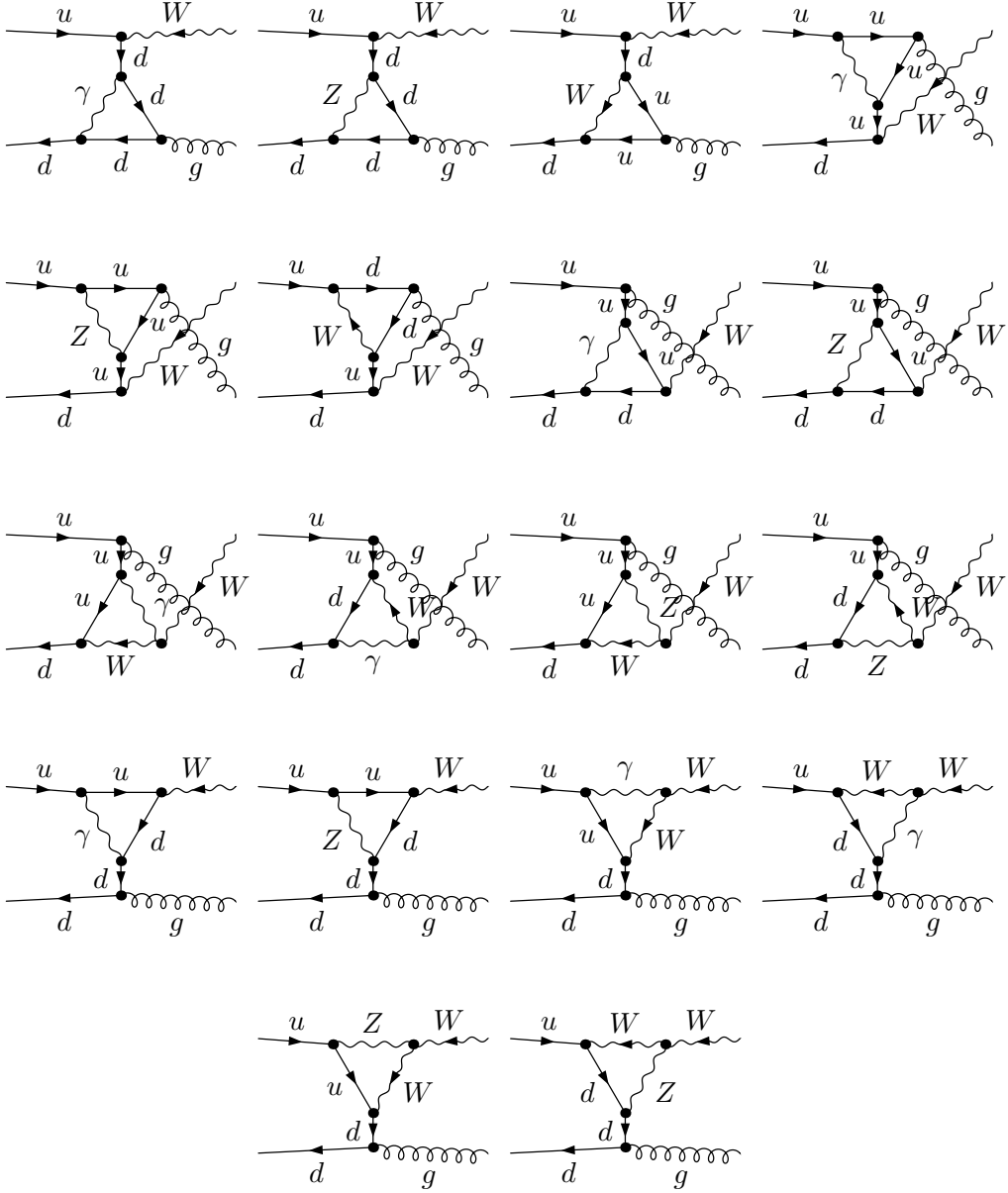}
\caption{$\mathcal{O}(\alpha)$ triangle diagrams of process~(\ref{eq:udg}).}
\label{triagudWg}
\end{center}
\end{figure}

\begin{figure}[ht]
\begin{center}
\input{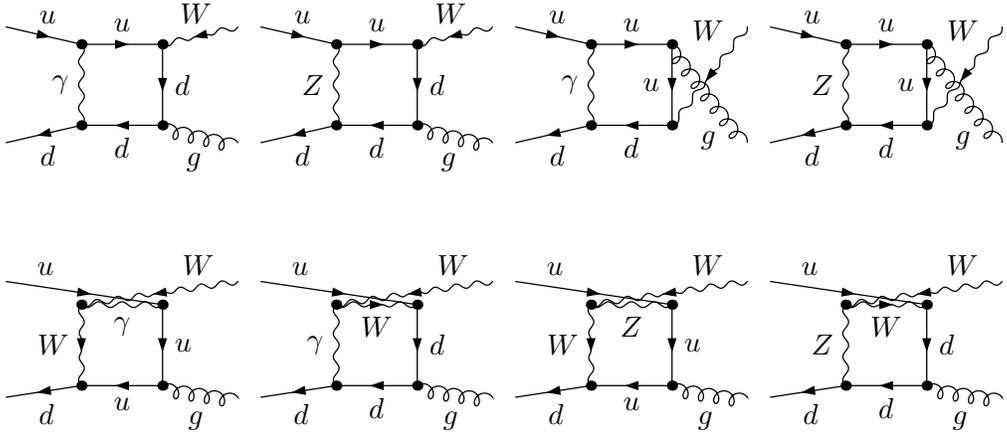}
\caption{$\mathcal{O}(\alpha)$ box diagrams of process~(\ref{eq:udg}).}
\label{boxesudWg}
\end{center}
\end{figure}

\begin{figure}[ht]
\begin{center}
\input{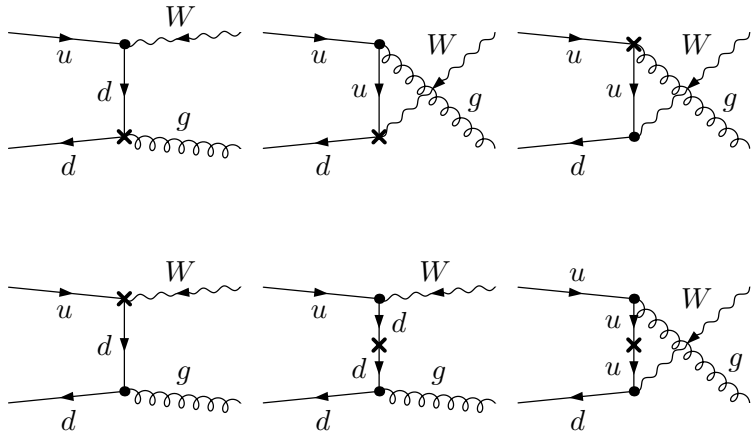}
\caption{$\mathcal{O}(\alpha)$ counterterm diagrams of process~(\ref{eq:udg}).}
\label{counterudwg}
\end{center}
\end{figure}

The virtual electroweak corrections of $\mathcal{O}(\alpha)$ to
processes~(\ref{eq:udg})--(\ref{eq:dgu}) arise from self-energy,
triangle, box, and counterterm diagrams.
They are shown for process~(\ref{eq:udg}) in
Figs.~\ref{selfenudWg}--\ref{counterudwg}, respectively.

Evaluating the transition matrix element
$\mathcal{T}_{\mathcal{O}(\alpha)}^{Wj}$ from these loop diagrams, we
encounter both ultraviolet (UV) and IR singularities, which need to be
regularised and removed.
As usual, we use dimensional regularisation, with $D=4-2\epsilon$ space-time
dimensions and 't~Hooft mass scale $\mu$, to extract the UV singularities as
single poles in $\epsilon$.
These are removed by renormalising the parameters and wave functions of
the LO transition matrix element $\mathcal{T}_0^{Wj}$, which leads to the
counterterm contribution (see Fig.~\ref{counterudwg}),
\begin{equation}
\mathcal{T}_\mathrm{CT}^{Wj}=\mathcal{T}_0^{Wj}\delta_\mathrm{CT}^{Wj}.
\end{equation}
Owing to the renormalisability of the SM, the UV singularities in
$\mathcal{T}_{\mathcal{O}(\alpha)}^{Wj}$ cancel, and the physical limit
$\epsilon\to0$ can be reached smoothly.

The electroweak on-shell renormalisation scheme uses the fine-structure
constant $\alpha$ defined in the Thomson limit and the physical particle
masses as basic parameters.
In order to avoid the appearance of large logarithms induced by the running of
$\alpha$ to the electroweak scale $M_W$ in
$\mathcal{T}_{\mathcal{O}(\alpha)}^{Wj}$, it is useful to replace $\alpha$ by
$G_F$ in the set of basic parameters, by substituting
\begin{equation}
G_F=\frac{\pi\alpha}{\sqrt{2}s_w^2w}\,\frac{1}{1-\Delta r},
\end{equation}
where $\Delta r$ \cite{Sirlin} contains those radiative corrections to the
muon lifetime which the SM introduces on top of those derived in the
QED-improved Fermi model.
In the electroweak on-shell scheme thus modified, we have
\begin{equation}
\delta_{\mathrm{CT}}^{Wj}=\delta Z_e-\frac{\delta s_w}{s_w}
+\frac{1}{2}\left(\delta Z_{u\overline{u}}^\mathrm{L}
+\delta Z_{d\overline{d}}^\mathrm{L}+\delta Z_W-\Delta r\right),
\end{equation}
where the renormalisation constants read \cite{Denner}
\begin{eqnarray}
\frac{\delta s_w}{s_w}&=&-\frac{1}{2}\,\frac{c_w^2}{s_w^2}\,
\re\left[\frac{\Sigma_{WW}^\mathrm{T}(M_W^2)}{M_W^2}
-\frac{\Sigma_{ZZ}^\mathrm{T}(M_Z^2)}{M_Z^2}\right],
\nonumber\\
\delta Z_e&=&\frac{1}{2}\,
\left.\frac{\partial\Sigma_{AA}^\mathrm{T}(q^2)}{\partial q^2}\right|_{q^2=0}
-\frac{s_w}{c_w}\,\frac{\Sigma_{AZ}^{\mathrm{T}}(0)}{M_Z^2},
\nonumber\\
\delta Z_W &=&-\re\left.\frac{\partial\Sigma_{WW}^\mathrm{T}(q^2)}
{\partial q^2}\right|_{q^2=M_W^2},
\nonumber\\ 
\delta Z_{q\overline{q}}^\mathrm{L}&=&-\re
\Sigma_{q\overline{q}}^\mathrm{L}(m_{q}^2)
-m_{q}^2\left.\frac{\partial}{\partial q^2}
\re\left[\Sigma_{q\overline{q}}^\mathrm{L}(q^2)
+\Sigma_{q\overline{q}}^\mathrm{R}(q^2)
+2\Sigma_{q\overline{q}}^\mathrm{S}(q^2)\right]\right|_{q^2=m_q^2}.
\label{renconew}
\end{eqnarray}
Here, $\Sigma_{WW}^\mathrm{T}$, $\Sigma_{ZZ}^\mathrm{T}$,
$\Sigma_{AA}^\mathrm{T}$, and $\Sigma_{AZ}^\mathrm{T}$ are the transverse
parts of the respective electroweak gauge-boson self-energies and mixing
amplitudes, $\Sigma_{q\overline{q}}^\mathrm{L}$,
$\Sigma_{q\overline{q}}^\mathrm{R}$, and $\Sigma_{q\overline{q}}^\mathrm{S}$
are the left-handed, right-handed, and scalar parts of the quark self-energy,
and $c_w^2=1-s_w^2$.

The IR singularities can be of soft or collinear type.
The loop diagrams involving virtual photons interchanged between external
lines are plagued by soft IR singularities.
Owing to the Bloch-Nordsieck theorem \cite{BloNo}, they cancel against similar
singularities arising from the real emission of soft photons, to be discussed
in Section~\ref{sec:udgp}.
The loop diagrams involving external quark or antiquark lines that split into
virtual photons and quarks generate collinear IR singularities.
Such singularities also arise from the real emission of collinear photons off
external quark or antiquark lines, as will be explained in
Section~\ref{sec:udgp}.
Thanks to the Kinoshita-Lee-Nauenberg theorem \cite{Kinoshita:1962ur},
collinear IR singularities from FSR are completely canceled in the sum of real
and virtual corrections provided that the final state is treated inclusively
enough.
On the other hand, collinear IR singularities from ISR survive and have to be
absorbed into the quark and antiquark PDFs.
For consistency, the splitting functions in the evolution equations of the PDFs
then need to be complemented by their $\mathcal{O}(\alpha)$ terms.
IR singularities also arise from the wave-function renormalisations in
Eq.~(\ref{renconew}).
We choose to regularise the IR singularities by assigning infinitesimal masses,
$\lambda$, $m_u$, and $m_d$, to the photon, the light up-type quarks, and the
down-type quarks, respectively.
This is convenient because the standard scalar one-loop integrals $C_0$ and
$D_0$ that emerge after the tensor reduction \cite{PaVe} are well established
for this regularisation prescription \cite{Beenakker:1988jr}.
Although the purely weak loop corrections are altogether devoid of IR
singularities, terms logarithmic in $m_u$ and $m_d$ are generated by the
tensor reduction.
However, these artificial IR singularities cancel among themselves.

We emphasise that, in the treatment of both the virtual and real corrections,
terms depending on $\lambda$, $m_u$, and $m_d$ are extracted analytically and
their cancellation is established manifestly, so that the expressions used for
the numerical analysis do not contain these IR regulators.

\subsection{Virtual QCD corrections to $W+\gamma$ production}
\label{sec:udp}

\begin{figure}[ht]
\begin{center}
\input{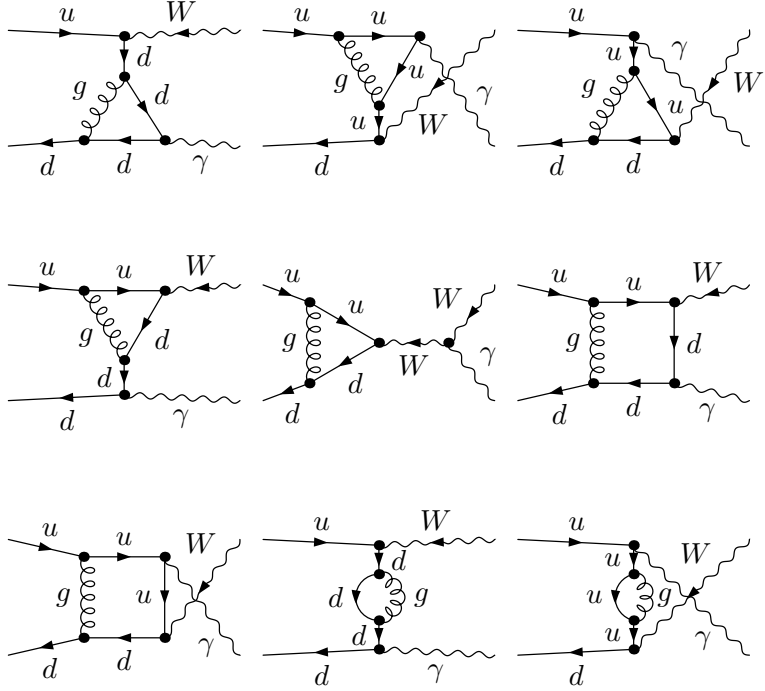}
\caption{$\mathcal{O}(\alpha_s)$ self-energy, triangle, and box diagrams of
process~(\ref{eq:udp}).}
\label{diagqcd}
\end{center}
\end{figure}

The virtual QCD corrections of $\mathcal{O}(\alpha_s)$ to
process~(\ref{eq:udp}) arise from the self-energy, triangle, and box diagrams
shown in Fig.~\ref{diagqcd} and the counterterm contribution,
\begin{equation}
\mathcal{T}_\mathrm{CT}^{W\gamma}=\mathcal{T}_0^{W\gamma}
\delta_\mathrm{CT}^{W\gamma}.
\end{equation}
The latter only receives contributions from the gluon-induced wave-function
renormalisation of the external quark lines,
\begin{equation}
\delta_\mathrm{CT}^{W\gamma}=\frac{1}{2}\left(\delta Z_{u\overline{u}}^g
+\delta Z_{d\overline{d}}^g\right),
\end{equation}
where
\begin{equation}
\delta Z_{q\overline{q}}^g=-\Sigma_{q\overline{q}}^{g,\mathrm{V}}(m_q^2)
-2 m_q^2\left.\frac{\partial}{\partial q^2}
\left[\Sigma_{q\overline{q}}^{g,\mathrm{V}}(q^2)
+\Sigma_{q\overline{q}}^{g,\mathrm{S}}(q^2)\right]\right|_{q^2=m_q^2}.
\end{equation}
Because parity is conserved within QCD, the quark self-energy has just one
vector part
$\Sigma_{q\overline{q}}^{g,\mathrm{V}}=\Sigma_{q\overline{q}}^{g,\mathrm{L}}=
\Sigma_{q\overline{q}}^{g,\mathrm{R}}$.
Up to terms that vanish in the limit $m_q\to0$, we have
\begin{equation}
\delta Z_{q\overline{q}}^g=
-\frac{\alpha_s C_F}{4\pi}\left[\frac{1}{\epsilon}-\gamma_E+\ln(4\pi)
-\ln\frac{m_q^2}{\mu^2}-2\ln\frac{m_q^2}{\lambda^2}+4\right]
+\mathcal{O}(\epsilon),
\end{equation}
where $C_F=(N_c^2-1)/(2N_c)=4/3$ for $N_c=3$ quark colours, $\gamma_E$ is the
Euler-Mascheroni constant, and $\lambda$ now represents an infinitesimal gluon
mass.

\subsection{Real corrections due to $W+j+\gamma$ production}
\label{sec:udgp}

\begin{figure}[ht]
\begin{center}
\input{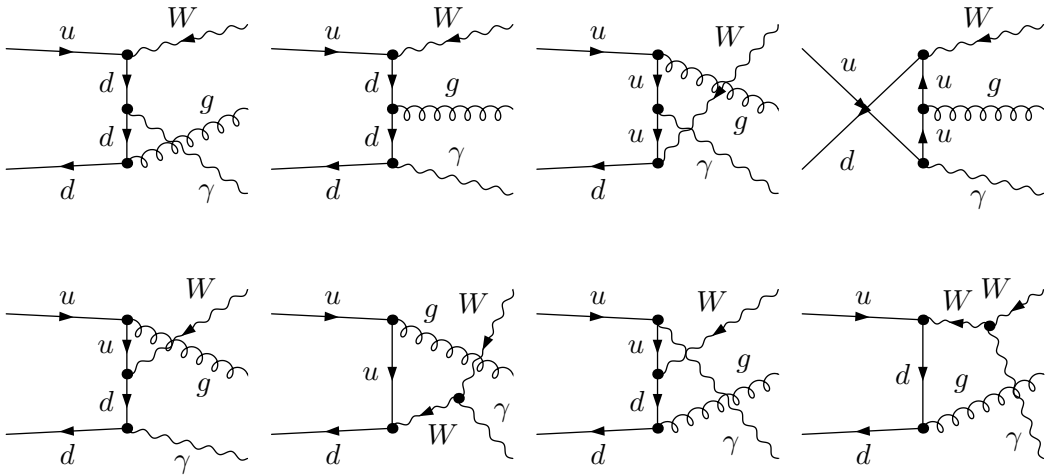}
\caption{Tree-level diagrams of process~(\ref{eq:udgp}).
The tree-level diagrams of processes~(\ref{eq:ugdp}) and (\ref{eq:dgup})
emerge through crossing.}
\label{fig:udgp}
\end{center}
\end{figure}

The tree-level diagrams for process~(\ref{eq:udgp}) are shown in
Fig.~\ref{fig:udgp}.
They contribute at the same time to the electromagnetic bremsstrahlung in
process~(\ref{eq:udg}) and to the QCD bremsstrah\-lung in
process~(\ref{eq:udp}), which complicates the treatment of the electroweak
corrections to $W+j$ associated production, as explained in the Introduction.
The diagrams contributing to the electromagnetic bremsstrahlung in
processes~(\ref{eq:ugd}) and (\ref{eq:dgu}) emerge from Fig.~\ref{fig:udgp} by
crossing the gluon with the $u$ and $\overline{d}$ quarks, respectively.

When the cross sections of processes~(\ref{eq:udgp})--(\ref{eq:dgup}) are
integrated over their three-particle phase spaces, one encounters IR
singularities of both soft and collinear types.
The former stem from the emission of soft photons and gluons and cancel
against similar contributions from the virtual corrections owing to the
Block-Nordsieck theorem \cite{BloNo}, as explained in Section~\ref{sec:udg}.
The latter arise when a massless gauge boson is collinearly emitted from an
external massless fermion line or when a massless gauge boson splits into two
collinear massless fermions.
Specifically, in process~(\ref{eq:udgp}), the photon or the gluon can be
emitted collinearly from the incoming $u_i$ and $\overline{d}_j$ quarks;
in process~(\ref{eq:ugdp}), the photon can be emitted collinearly from the
incoming $u_i$ quark or the outgoing $d_j$ quark, and the gluon can split
into a collinear $d_j\overline{d}_j$ quark pair;
and in process~(\ref{eq:dgup}), the photon can be emitted collinearly from the
incoming $\overline{d}_j$ quark or the outgoing $\overline{u}_i$ quark, and
the gluon can split into a collinear $u_i\overline{u}_i$ quark pair.
As already mentioned in Section~\ref{sec:udg}, the collinear IR singularities
from FSR are canceled by the virtual corrections according to the
Kinoshita-Lee-Nauenberg theorem \cite{Kinoshita:1962ur} if the considered
process is treated inclusively enough.
By contrast, those from ISR survive and have to be absorbed into the PDFs.

Due to the minimum-$p_T$ cut, the photon and the gluon cannot be soft
simultaneously because one of them has to balance the transverse momentum of
the $W$ boson.
By the same token, there can only be one collinear situation at a time.
However, soft and collinear singularities do overlap, and care needs to be
exercised to avoid double counting.

For consistency, also the IR singularities in the real corrections need to be
regularised by the photon and gluon mass $\lambda$ and the light-quark masses
$m_u$ and $m_d$ introduced in Sections~\ref{sec:udg} and \ref{sec:udp}.
As already mentioned in Section~\ref{sec:udg}, their cancellation is achieved
analytically, so that the expressions underlying the numerical analysis are
free of them.

As in Ref.~\cite{Baur:1998kt,Diener,DieDitHol}, we employ the method of phase
space slicing \cite{Fabricius:1981sx} to separate the soft and collinear
regions of the phase space from the one where the momenta are hard and
non-collinear, so that the partonic cross section can be written as
\begin{equation}
\mathrm{d}\hat\sigma^{Wj\gamma}=\hat\sigma_\mathrm{soft}^{Wj\gamma}
+\mathrm{d}\hat\sigma_\mathrm{coll}^{Wj\gamma}
+\mathrm{d}\hat\sigma_\mathrm{hard}^{Wj\gamma}.
\end{equation}
For definiteness, let us assume that parton $d$ in process~(\ref{eq:subproc})
is the soft or collinearly emitted one and that partons $a$ and $c$ are the
ones emitting ISR and FSR, respectively.
In the notation introduced in Section~\ref{sec:tpfs}, the soft regions of
phase space are then defined by $\lambda<p_d^0<\Delta E\ll(s-w)/(2\sqrt{s})$,
the collinear ones for ISR and FSR by $\vartheta<\Delta\vartheta\ll\pi$ and
$\psi<\Delta\psi\ll\pi$, respectively, and $p_d^0>\Delta E$, and the hard and
non-collinear one by the rest.
In Sections~\ref{sec:soft} and \ref{sec:coll}, we explain how to evaluate
$\mathrm{d}\hat\sigma_\mathrm{soft}^{Wj\gamma}$ and
$\mathrm{d}\hat\sigma_\mathrm{coll}^{Wj\gamma}$ analytically using appropriate
approximations.
On the other hand, $\mathrm{d}\hat\sigma_\mathrm{hard}^{Wj\gamma}$ can
straightforwardly be evaluated numerically with high precision
\cite{Monyonko:1985iq}.
Since $\Delta E$ is to be measured in units of $\sqrt{s}/2$, we define
$\delta_s=2\Delta E/\sqrt{s}$.
The demarcation parameters $\delta_s$, $\Delta\vartheta$, and $\Delta\psi$
must be chosen judiciously.
If the are too small, then the numerical phase-space integration performed for
$\mathrm{d}\hat\sigma_\mathrm{hard}^{Wj\gamma}$ becomes unstable; if they are
too large, the approximations adopted for
$\mathrm{d}\hat\sigma_\mathrm{soft}^{Wj\gamma}$ and
$\mathrm{d}\hat\sigma_\mathrm{coll}^{Wj\gamma}$ become crude.
In practice, one varies $\delta_s$, $\Delta\vartheta$, and $\Delta\psi$ to
find the respective stability regions.
For the problem considered here, this is easily achieved. 

\subsubsection{Soft singularities}
\label{sec:soft}

In the soft phase space regions, $\mathcal{T}^{ab\to Wcd}$ factorises into
$\mathcal{T}^{ab\to Wc}$ times an eikonal factor that depends on $\vec{p}_d$.
Squaring $\mathcal{T}^{ab\to Wcd}$, performing the spin and colour sums, and
integrating over $\vec{p}_d$ with the constraint $\lambda<p_d^0<\Delta E$, one
has \cite{Denner,ThooftVelt}
\begin{equation}
\mathrm{d}\hat\sigma_\mathrm{soft}^{ab\to Wcd}(\lambda,\Delta E)
=\delta_\mathrm{soft}^{ab\to Wcd}(\lambda,\Delta E)
\mathrm{d}\hat\sigma^{ab\to Wc}.
\end{equation}
In the case of soft electromagnetic and QCD bremsstrahlung in
process~(\ref{eq:udgp}), we then obtain
\begin{eqnarray}
\delta_\mathrm{soft}^{u\overline{d}\to W^+g\gamma}(\lambda,\Delta E)&=&
-\frac{\alpha}{2\pi}\left(Q_u^2\delta_{uu}+Q_d^2\delta_{dd}+\delta_{WW}
+2Q_uQ_d\delta_{ud}+2Q_u\delta_{uW}+2Q_d\delta_{dW}\right),
\nonumber\\
\delta_\mathrm{soft}^{u\overline{d}\to W^+\gamma g}(\lambda,\Delta E)&=&
-\frac{\alpha_sC_F}{2\pi}(\delta_{uu}+\delta_{dd}+2\delta_{ud}),
\end{eqnarray}
where $Q_u=2/3$ and $Q_d=-1/3$ are the fractional electric charges of the $u$
and $d$ quarks, respectively, and 
\begin{eqnarray}
\delta_{uu}&=&\ln\frac{4(\Delta E)^2}{\lambda^2}+\ln\frac{m_u^2}{s},
\nonumber\\
\delta_{dd}&=&\left.\delta_{uu}\right|_{m_u\leftrightarrow m_d},
\nonumber\\
\delta_{ud}&=&\frac{1}{2}\ln\frac{4(\Delta E)^2}{\lambda^2}
\ln\frac{m_u^2m_d^2}{s^2}
+\frac{1}{4}\left(\ln^2\frac{m_u^2}{s}+\ln^2\frac{m_d^2}{s}\right)
+\frac{\pi^2}{3},
\nonumber\\
\delta_{WW}&=&\ln\frac{4(\Delta E)^2}{\lambda^2}+\frac{s+w}{s-w}\ln\frac{w}{s},
\nonumber\\
\delta_{uW}&=&\frac{1}{2}\ln\frac{4(\Delta E)^2}{\lambda^2}
\ln\frac{wm_u^2}{(w-t)^2}
+\frac{1}{4}\left(\ln^2\frac{m_u^2}{s}+\ln^2\frac{w}{s}\right)
\nonumber\\
&&{}+\li\left(\frac{-t}{w-t}\right)+\li\left(\frac{-u}{w-t}\right)
+\frac{\pi^2}{6},
\nonumber\\
\delta_{dW}&=&-\left.\delta_{uW}
\right|_{t\leftrightarrow u,\,m_u\leftrightarrow m_d}.
\label{eq:soft}
\end{eqnarray}
Here, $\li(x)=-\int_0^1\mathrm{d}t\,\ln(1-tx)/t$ is the dilogarithm, and terms
that vanish for $m_u=m_d=0$ have been omitted.

Furthermore, we find the soft-photon correction factor for
process~(\ref{eq:ugdp}) to be
\begin{equation}
\delta_\mathrm{soft}^{ug\to W^+d\gamma}(\lambda,\Delta E)=
-\frac{\alpha}{2\pi}\left(Q_u^2\delta_{uu}+Q_d^2\delta_{dd}+\delta_{WW}
+2Q_uQ_d\tilde\delta_{ud}+2Q_u\delta_{uW}+2Q_d\tilde\delta_{dW}\right),
\end{equation}
in which two terms of Eq.~(\ref{eq:soft}) are modified to be
\begin{eqnarray}
\tilde\delta_{ud}&=&\frac{1}{2}\ln\frac{4(\Delta E)^2}{\lambda^2}
\ln\frac{m_u^2m_d^2}{u^2}
+\frac{1}{4}\left[\ln^2\frac{m_u^2}{s}+\ln^2\frac{sm_d^2}{(s-w)^2}\right]
+\li\left(-\frac{t}{u}\right)+\frac{\pi^2}{3},
\nonumber\\
\tilde\delta_{dW}&=&-\frac{1}{2}\ln\frac{4(\Delta E)^2}{\lambda^2}
\ln\frac{wm_u^2}{(s-w)^2}
-\frac{1}{4}\left[\ln^2\frac{w}{s}+\ln^2\frac{sm_d^2}{(s-w)^2}\right]
-\li\left(1-\frac{w}{s}\right)-\frac{\pi^2}{6}.
\nonumber\\
&&
\end{eqnarray}

Finally, the soft-photon correction factor for process~(\ref{eq:dgup}) emerges
from the one of process~(\ref{eq:ugdp}) through the simple replacement
\begin{equation}
\delta_\mathrm{soft}^{\overline{d}g\to W^+\overline{u}\gamma}(\lambda,\Delta E)
=\left.\delta_\mathrm{soft}^{ug\to W^+d\gamma}
(\lambda,\Delta E)\right|_{m_u\leftrightarrow m_d}.
\end{equation}

\subsubsection{Collinear singularities}
\label{sec:coll}

As explained in Section~\ref{sec:udgp}, collinear singularities arise from
three sources:
(1) the emission of a photon or gluon from an incoming quark or antiquark;
(2) the splitting of an incoming gluon into a quark-antiquark pair; and
(3) the emission of a photon from an outgoing quark or antiquark.
The resulting contributions to $\mathrm{d}\hat\sigma_\mathrm{coll}^{Wj\gamma}$
all factorise into the respective LO cross sections without radiation and
appropriate collinear radiator functions \cite{DittmDoktor,Kleiss}.
In the case of ISR, this also involves a convolution with respect to the
fraction $z$ of four-momentum that the emitting parton passes on to the one
that enters the hard interaction.

Let parton $a$ in process~(\ref{eq:subproc}) be the emitting quark $q$ and
parton $d$ the emitted photon or gluon.
Then we have \cite{DittmDoktor,Kleiss}
\begin{equation}
\mathrm{d}\hat\sigma_\mathrm{coll}^{qb\to Wc\{\gamma,g\}}(m_q,\Delta\vartheta)
=\frac{\left\{\alpha Q_q^2,\alpha_s C_F\right\}}{2\pi}
\int_{z_0}^{1-\delta_s}\mathrm{d}z\,R_q^\mathrm{IS}(m_q,\Delta\vartheta,z)
\left.\mathrm{d}\hat\sigma_0^{qb\to Wc}\right|_{p_q\to zp_q},
\end{equation}
where $\delta_s$ is introduced to exclude a slice of phase space that is both
soft and collinear and is already included in
$\hat\sigma_\mathrm{soft}^{Wj\gamma}$, $z_0=\tau_0S/s$, with $\tau_0$ being
defined in Eq.~(\ref{eq:tau}), and
\begin{equation}
R_q^{\mathrm{IS}}(m_q,\Delta\vartheta,z)=P_{q\to q}(z)
\left[\ln\frac{s(\Delta\vartheta)^2}{4m_q^2}-\frac{2z}{1+z^2}\right],
\end{equation}
with
\begin{equation}
P_{q\to q}(z)=\frac{1+z^2}{1-z}
\label{eq:pqq}
\end{equation}
being the LO $q\to q$ splitting function \cite{Altarelli:1977zs}.
This result readily carries over to the case when parton $a$ is an antiquark
$\overline{q}$.
Note that the c.m.\ frame is boosted along the beam axis by the collinear
emission of the photon or gluon.

Now, let parton $a$ in process~(\ref{eq:subproc}) be a gluon that splits into
a $q\overline{q}$ pair, with $q$ being outgoing and $\overline{q}$ entering
the residual hard scattering.
Then we have \cite{DieDitHol}
\begin{equation}
\mathrm{d}\hat\sigma_\mathrm{coll}^{gb\to Wcq}(m_q,\Delta\vartheta)
=\frac{\alpha_s T_F}{2\pi}
\int_{z_0}^1\mathrm{d}z\,R_g^\mathrm{IS}(m_q,\Delta\vartheta,z)
\left.\mathrm{d}\hat\sigma_0^{\overline{q}b\to Wc}
\right|_{p_{\overline{q}}\to zp_{\overline{q}}},
\end{equation}
where $T_F=1/2$ and
\begin{equation}
R_g^{\mathrm{IS}}(m_q,\Delta\vartheta,z)=
P_{g\to q}(z)\ln\frac{s(1-z)^2(\Delta\vartheta)^2}{4m_q^2}+2z(1-z),
\end{equation}
with
\begin{equation}
P_{g\to q}(z)=z^2+(1-z)^2
\end{equation}
being the LO $g\to q$ splitting function.
This result readily carries over to the case when parton $d$ is an antiquark
$\overline{q}$.

Finally, let parton $c$ in process~(\ref{eq:subproc}) be the emitting quark
$q$ and parton $d$ the emitted photon.
Then we have \cite{DittmDoktor,Kleiss}
\begin{equation}
\mathrm{d}\hat\sigma_\mathrm{coll}^{ab\to Wq\gamma}(m_q,\Delta\psi)
=\frac{\alpha Q_q^2}{2\pi}
\int_0^{1-\tilde\delta_s}\mathrm{d}z\,R_q^\mathrm{FS}(m_q,\Delta\psi,z)
\mathrm{d}\hat\sigma_0^{ab\to Wq},
\label{eq:fsr}
\end{equation}
where $\tilde\delta_s=s\delta_s/(s-w)$ is again to avoid double counting of
phase space regions that are both soft and collinear, and
\begin{equation}
R_q^{\mathrm{FS}}(m_q,\Delta\psi,z)=P_{q\to q}(z)
\left[\ln\frac{(s-w)^2(\Delta\psi)^2}{4sm_q^2}+2\ln z-\frac{2z}{1+z^2}\right],
\end{equation}
with $P_{q\to q}$ given in Eq.~(\ref{eq:pqq}).
This result readily carries over to the case when parton $c$ is an antiquark
$\overline{q}$.
The integral in Eq.~(\ref{eq:fsr}) is not a convolution and can easily be
carried out, yielding
\begin{equation}
\int_0^{1-\tilde\delta_s}\mathrm{d}z\,R_q^\mathrm{FS}(m_q,\Delta\psi,z)
=\left(-2\ln\tilde\delta_s-\frac{3}{2}\right)
\ln\frac{(s-w)^2(\Delta\psi)^2}{4sm_q^2}+2\ln\tilde\delta_s-\frac{2}{3}\pi
+\frac{9}{2}.
\end{equation}

In order to obtain $\mathrm{d}\hat\sigma_\mathrm{coll}^{Wj\gamma}$ for one of
the processes~(\ref{eq:udgp})--(\ref{eq:dgup}), all possible collinear
emissions must be taken into account one by one.

While the collinear IR singularities from FSR cancel upon combination with the
virtual corrections by the Kinoshita-Lee-Nauenberg theorem
\cite{Kinoshita:1962ur}, those from ISR survive.
Since their form is universal, they can be factorised and absorbed into the
PDFs \cite{Collins:1989gx}.
Adopting the modified minimal-subtraction ($\overline{\mathrm{MS}}$)
factorisation scheme both for the collinear singularities of relative orders
$\mathcal{O}(\alpha)$ and $\mathcal{O}(\alpha_s)$, this is achieved by
modifying the PDF of quark $q$ inside hadron $A$ as
\begin{eqnarray}
\lefteqn{f_{q/A}(x,M^2)\to\tilde{f}_{q/A}(x,M^2)=
f_{q/A}(x,M^2)\left\{1-\frac{\alpha Q_q^2+\alpha_sC_F}{\pi}
\left[\left(\ln\delta_s+\frac{3}{4}\right)\ln\frac{M^2}{m_q^2}
\right.\right.}
\nonumber\\
&&{}-\left.\left.\ln^2\delta_s-\ln\delta_s+1\vphantom{\frac{M^2}{m_q^2}}
\right]\right\}
-\int_x^{1-\delta_s}\frac{\mathrm{d}z}{z}\,f_{q/A}\left(\frac{x}{z},M^2\right)
\frac{\alpha Q_q^2+\alpha_sC_F}{2\pi}P_{q\to q}(z)
\nonumber\\
&&{}\times
\left[\ln\frac{M^2}{(1-z)^2m_q^2}-1\right]
-\int_{x}^1\frac{\mathrm{d}z}{z}\,f_{g/A}\left(\frac{x}{z},M^2\right)
\frac{\alpha_s T_F}{2\pi}P_{g\to q}(z)\ln\frac{M^2}{m_q^2},\qquad\qquad\qquad
\end{eqnarray}
where $M$ is the factorisation mass scale, which separates the perturbative
and non-perturbative parts of the hadronic cross section.

\section{Numerical results}
\label{sec:four}

We are now in a position to present our numerical results.
We start by specifying our choice of input.
We adopt the values $G_F=1.6637\times10^{-5}$~GeV$^{-2}$, $M_W=80.403$~GeV,
$M_Z=91.1876$~GeV, and $m_t=174.2$~GeV recently quoted by the Particle Data
Group \cite{pdg}, take the other $n_f=5$ quarks to be massless partons, and
assume $M_H=120$~GeV, which is presently compatible with the direct search
limits and the bounds from the electroweak precision tests \cite{pdg}.
We take the absolute values of the CKM matrix elements to be \cite{pdg}
\begin{equation}
\begin{array}{lll}
|V_{ud}|=0.9377, & |V_{us}|=0.2257, & |V_{cd}|=0.230, \\
|V_{cs}|=0.957, & |V_{cb}|=41.6\times10^{-3}, & |V_{ub}|=4.31\times 10^{-3}.
\end{array}
\end{equation} 
Since we are working at LO in QCD, we employ the one-loop formula for
$\alpha_s^{(n_f)}(\mu)$.
We use the LO proton PDF set \verb/CTEQ6L1/ by the Coordinated
Theoretical-Experimental Project on QCD (CTEQ) \cite{CTEQ}, with
$\Lambda_\mathrm{QCD}^{(5)}=165$~MeV.
In the case of photoproduction, we add the photon spectra for elastic
\cite{kni} and inelastic \cite{gsv,Martin:2004dh} scattering elaborated in
the Weizs\"acker-Williams approximation.
In the latter case, we use the more recent set by Martin, Roberts, Stirling,
and Thorne (\verb/MRSTQED04/) \cite{Martin:2004dh} as our default, and the set
by Gl\"uck, Stratmann, and Vogelsang (\verb/GSV/) \cite{gsv} to assess the
theoretical uncertainty from this source.
We choose the renormalisation and factorisation scales to be
$\mu=M=\xi m_T^\mathrm{cut}$, where
$m_T^\mathrm{cut}=\sqrt{(p_T^\mathrm{cut})^2+M_W^2}$ is the minimum
transverse mass of the produced $W$ boson and $\xi$ is introduced to estimate
the theoretical uncertainty.
Unless otherwise stated, we use the default value $\xi=1$.

We consider the total cross sections of $p\overline{p}\to W^\pm+X$ at the
Tevatron (run~II) with $\sqrt{S}=1.96$~TeV and $pp\to W^\pm+X$ at the LHC with
$\sqrt{S}=14$~TeV as functions of $p_T^\mathrm{cut}$.
By numerically differentiating the latter with respect to $p_T^\mathrm{cut}$,
we also obtain the corresponding $p_T$ distributions as
$\mathrm{d}\sigma/\mathrm{d}p_T=-\left.\mathrm{d}\sigma(p_T^\mathrm{cut})/
\mathrm{d}p_T^\mathrm{cut}\right|_{p_T^\mathrm{cut}=p_T}$.
Owing to the baryon symmetry of the initial state, the results for $W^+$ and
$W^-$ bosons are identical at the Tevatron, and it is sufficient to study one
of them.
By contrast, $W^+$-boson production is favoured at the LHC because the proton
most frequently interacts via a $u$ quark.
Therefore, it is necessary to study the production of $W^+$ and $W^-$ bosons
separately at the LHC.
We compare the contributions of four different orders:
(1) the LO contribution of $\mathcal{O}(\alpha\alpha_s)$ from
processes~(\ref{eq:udg})--(\ref{eq:dgu}), where the system $X$ accompanying
the $W$ boson contains a hadron jet;
(2) the LO contribution of $\mathcal{O}(\alpha^2)$ from process~(\ref{eq:udp}),
where $X$ contains a prompt photon;
(3) the NLO contribution of $\mathcal{O}(\alpha^2\alpha_s)$ comprising
processes~(\ref{eq:udg})--(\ref{eq:udp}) at one loop as well as
processes~(\ref{eq:udgp})--(\ref{eq:dgup}) at tree level, where $X$ contains
a hadron jet, a prompt photon, or both; and
(4) the LO contributions of $\mathcal{O}(\alpha^3)$ from
processes~(\ref{eq:upd}) and (\ref{eq:dpu}) via direct photoproduction and
from processes~(\ref{eq:udg})--(\ref{eq:dgu}) via resolved photoproduction,
where $X$ contains a hadron jet and, in the case of elastic photoproduction,
also the scattered proton or antiproton.
Since we consider inclusive one-particle production, we do not use any
information on the composition of $X$, i.e.\ we include all possibilities.
In the following, we regard the sum of contributions (1) and (2) as LO and
sum of contributions (1)--(4) as NLO unless the perturbative orders are
explicitly specified in terms of coupling constants.
We thus define the correction factor $K$ to be the NLO to LO ratio with this
understanding.

\begin{figure}[ht]
\begin{center}
\begin{tabular}{ll}
\parbox{0.45\textwidth}{
\includegraphics[bb=170 460 425 705,width=0.45\textwidth]%
{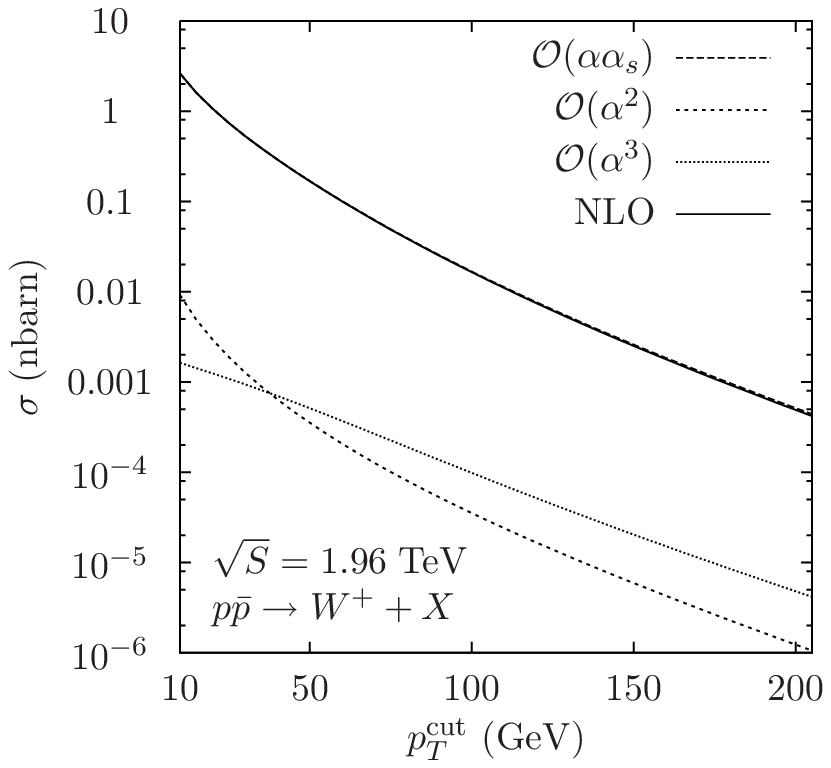}
}
&
\parbox{0.45\textwidth}{
\includegraphics[bb=170 460 425 705,width=0.45\textwidth]%
{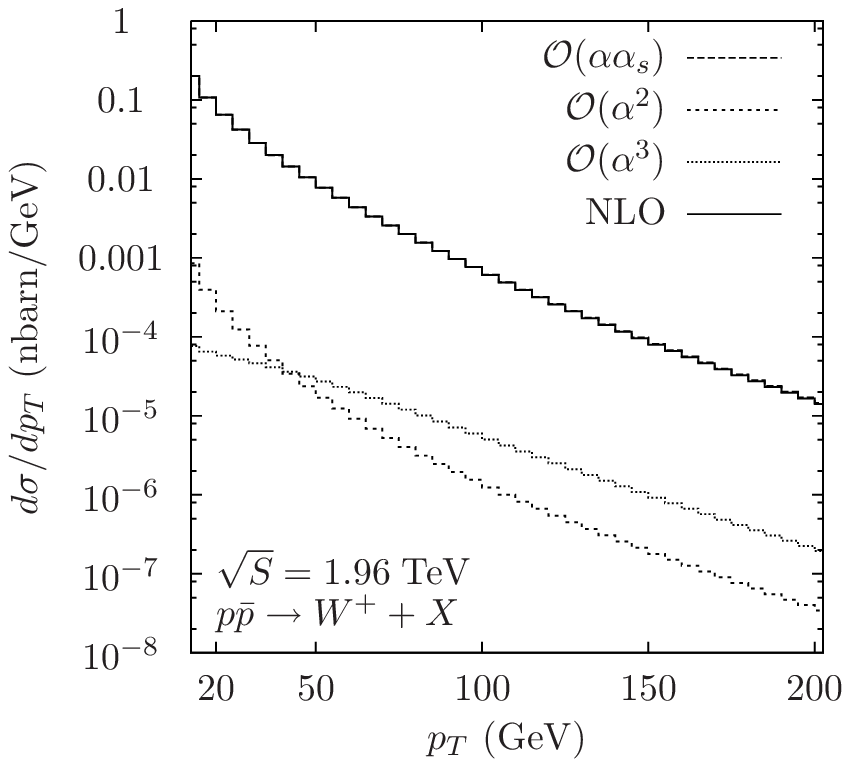}
}
\\
(a) & (b)
\end{tabular}
\caption{(a) Total cross section as a function of $p_T^\mathrm{cut}$ and (b)
$p_T$ distribution of $p\overline{p}\to W^++X$ for $\sqrt{S}=1.96$~TeV
(Tevatron run~II).
The NLO results are compared with those of orders
$\mathcal{O}(\alpha\alpha_s)$, $\mathcal{O}(\alpha^2)$, and
$\mathcal{O}(\alpha^3)$ via photoproduction.}
\label{fig:Tev}
\end{center}
\end{figure}

\begin{figure}[ht]
\begin{center}
\begin{tabular}{ll}
\parbox{0.45\textwidth}{
\includegraphics[bb=170 430 425 705,width=0.45\textwidth]%
{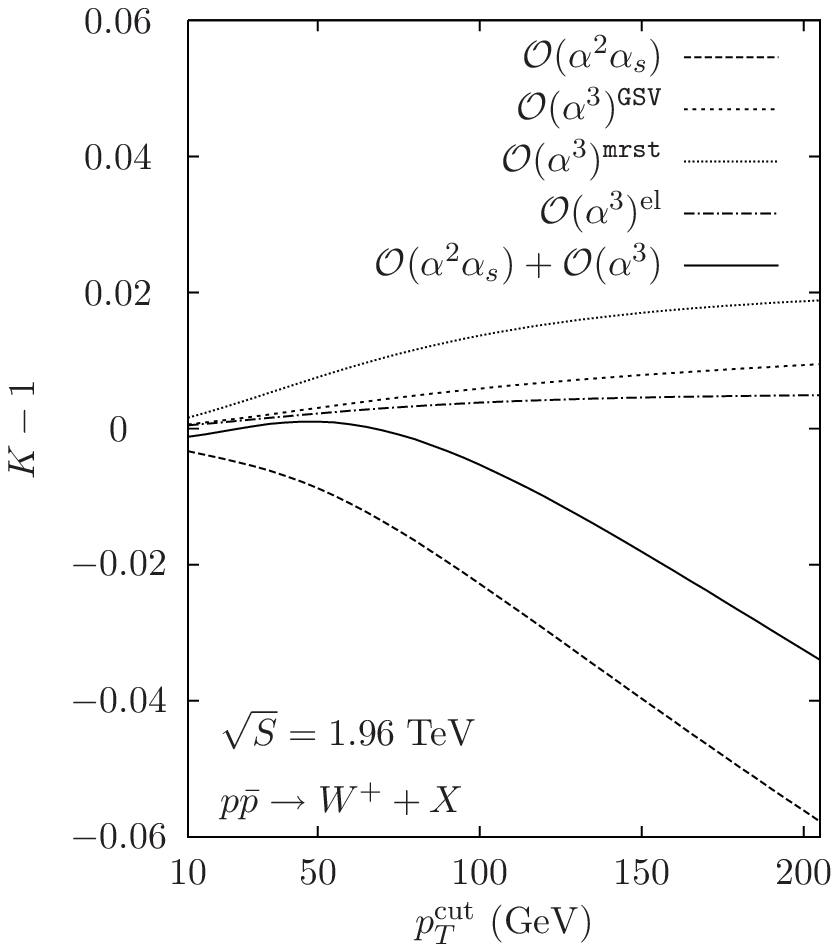} 
}
&
\parbox{0.45\textwidth}{
\includegraphics[bb=170 430 425 705,width=0.45\textwidth]%
{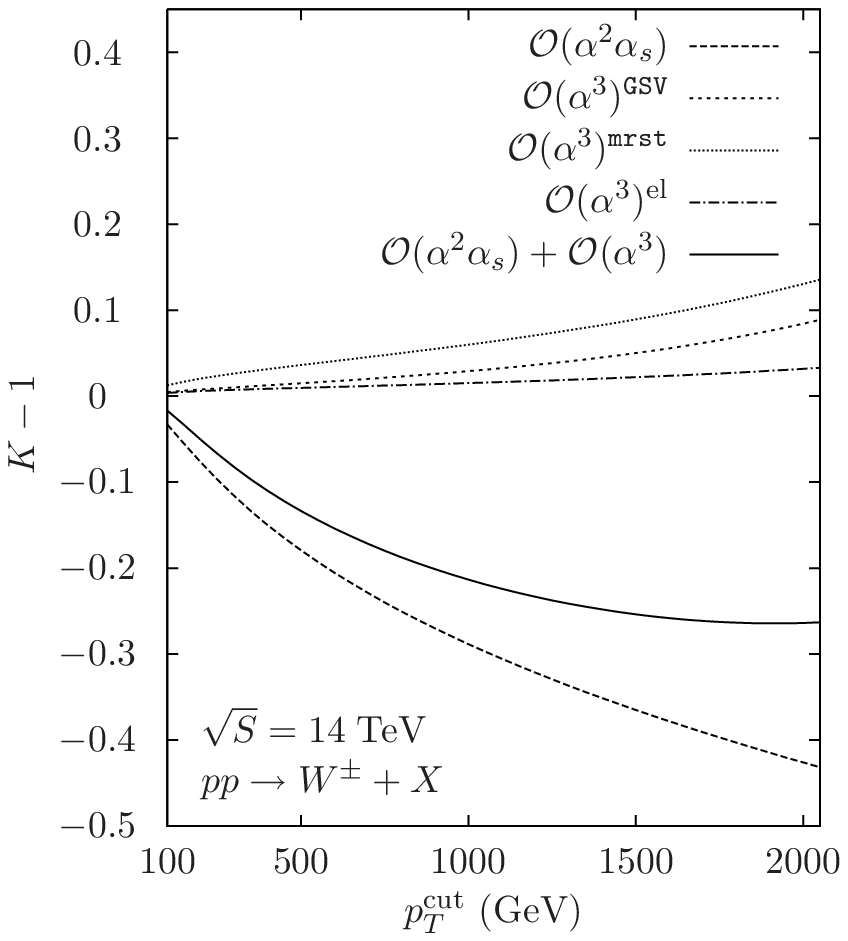}
}
\\
(a) & (b)
\end{tabular}
\caption{NLO corrections ($K-1$), with and without the photoproduction
contributions, to the total cross sections of (a)
$p\overline{p}\to W^++X$ for $\sqrt{S}=1.96$~TeV (Tevatron run~II) and of (b)
$pp\to W^\pm+X$ for $\sqrt{S}=14$~TeV (LHC) as functions $p_T^\mathrm{cut}$.
For comparison, also the contributions due to elastic and inelastic
photoproduction normalised to the LO results are shown.
In the latter case, the evaluation is also performed with the {\tt GSV} PDFs.}
\label{fig:K}
\end{center}
\end{figure}

\begin{figure}[ht]
\begin{center}
\begin{tabular}{ll}
\parbox{0.45\textwidth}{
\includegraphics[bb=170 460 425 705,width=0.45\textwidth]%
{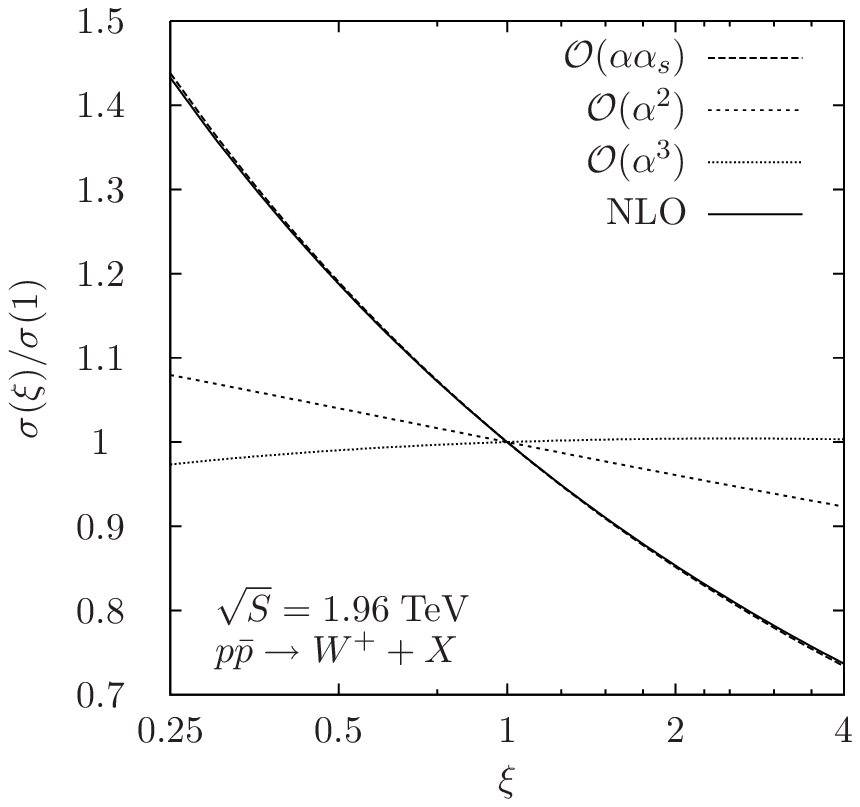}
}
&
\parbox{0.45\textwidth}{
\includegraphics[bb=170 460 425 705,width=0.45\textwidth]%
{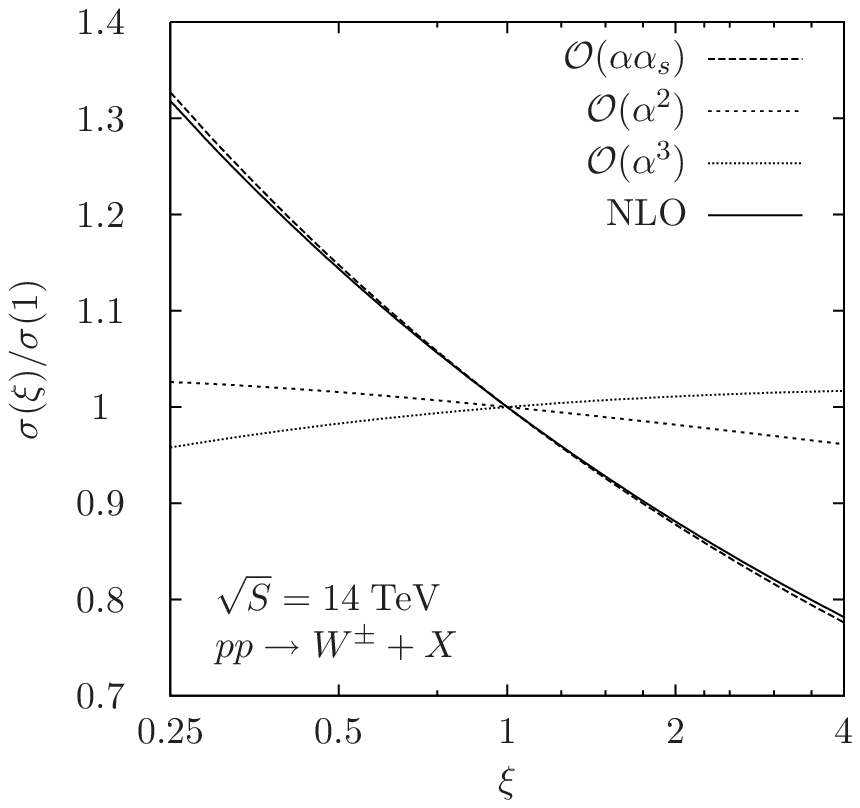}
}
\\
(a) & (b)
\end{tabular}
\caption{Total cross sections of (a) $p\overline{p}\to W^++X$ for
$\sqrt{S}=1.96$~TeV and $p_T^\mathrm{cut}=20$~GeV (Tevatron run~II) and of (b)
$pp\to W^\pm+X$ for $\sqrt{S}=14$~TeV and $p_T^\mathrm{cut}=200$~GeV (LHC) as
functions of $\xi$ normalised to their default values for $\xi=1$.
The NLO results are compared with those of orders
$\mathcal{O}(\alpha\alpha_s)$, $\mathcal{O}(\alpha^2)$, and
$\mathcal{O}(\alpha^3)$ via photoproduction.}
\label{fig:xi}
\end{center}
\end{figure}

\begin{figure}[ht]
\begin{center}
\begin{tabular}{ll}
\parbox{0.45\textwidth}{
\includegraphics[bb=170 460 425 705,width=0.45\textwidth]%
{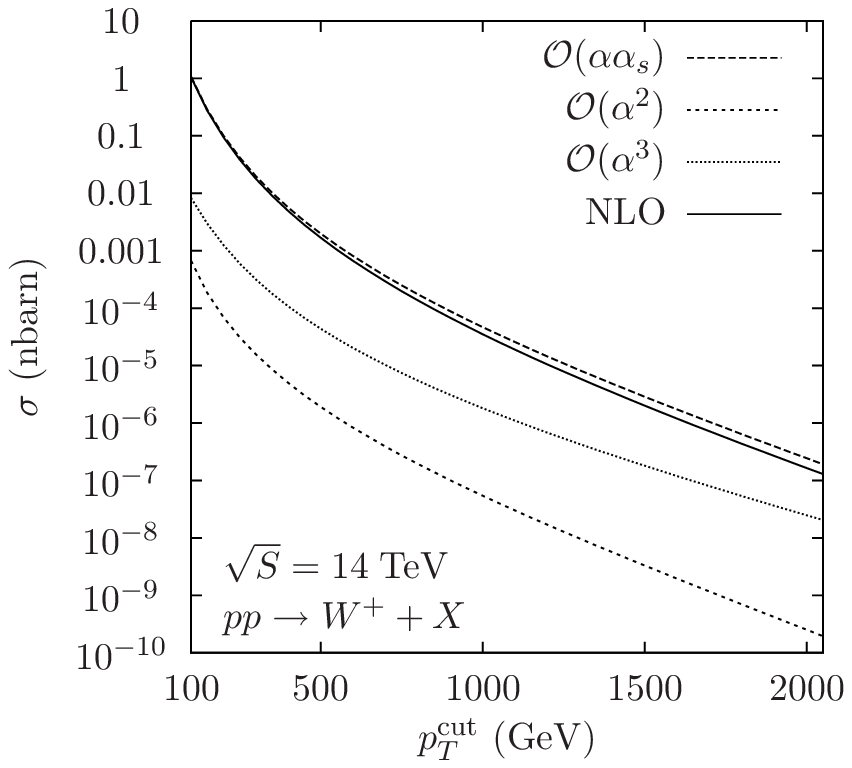}
}
&
\parbox{0.45\textwidth}{
\includegraphics[bb=170 460 425 705,width=0.45\textwidth]%
{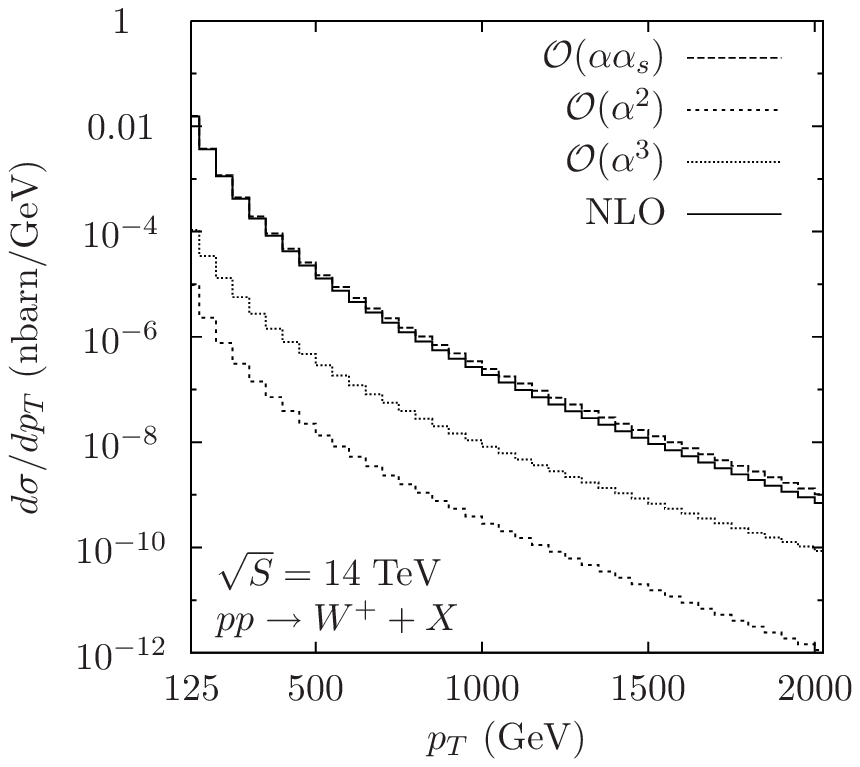}
}
\\
(a) & (b)
\end{tabular}
\caption{(a) Total cross section as a function of $p_T^\mathrm{cut}$ and (b)
$p_T$ distribution of $pp\to W^++X$ for $\sqrt{S}=14$~TeV (LHC).
The NLO results are compared with those of orders
$\mathcal{O}(\alpha\alpha_s)$, $\mathcal{O}(\alpha^2)$, and
$\mathcal{O}(\alpha^3)$ via photoproduction.}
\label{fig:LHCp}
\end{center}
\end{figure}

\begin{figure}[ht]
\begin{center}
\begin{tabular}{ll}
\parbox{0.45\textwidth}{
\includegraphics[bb=170 460 425 705,width=0.45\textwidth]%
{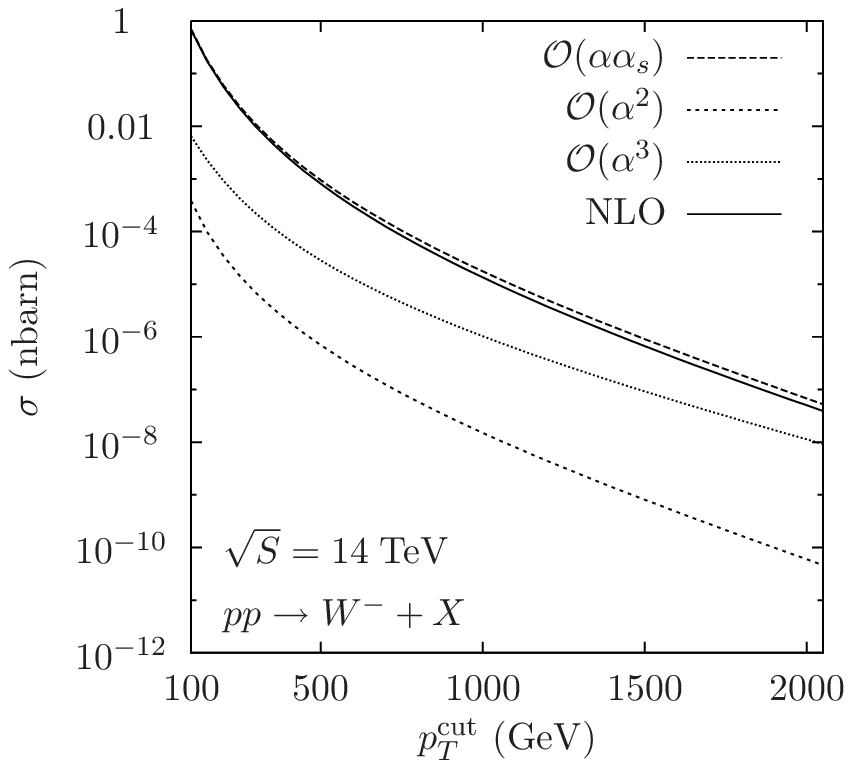}
}
&
\parbox{0.45\textwidth}{
\includegraphics[bb=170 460 425 705,width=0.45\textwidth]%
{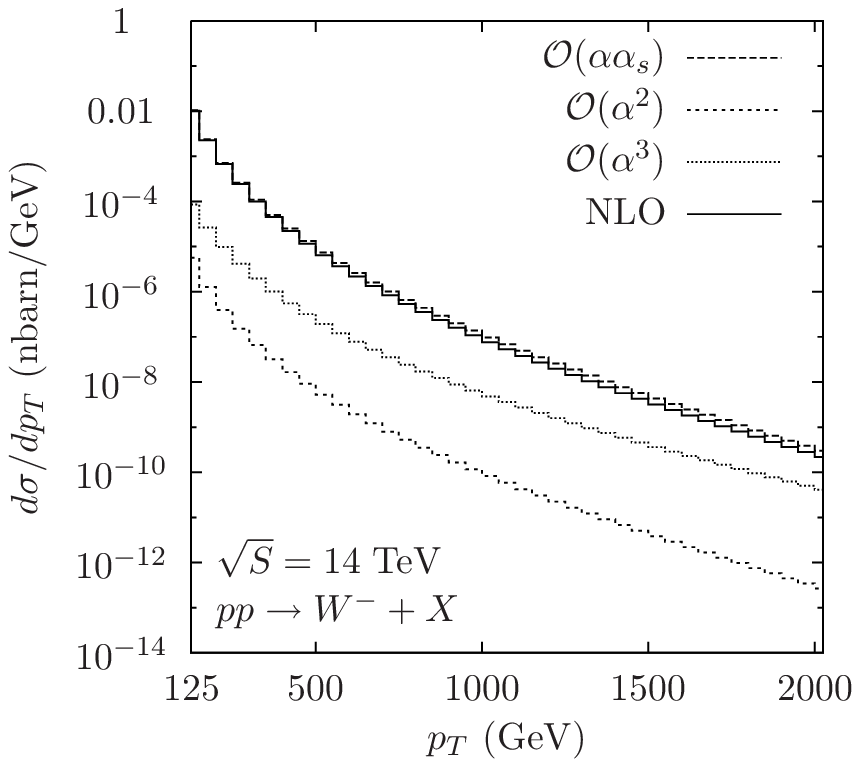}
}
\\
(a) & (b)
\end{tabular}
\caption{(a) Total cross section as a function of $p_T^\mathrm{cut}$ and (b)
$p_T$ distribution of $pp\to W^-+X$ for $\sqrt{S}=14$~TeV (LHC).
The NLO results are compared with those of orders
$\mathcal{O}(\alpha\alpha_s)$, $\mathcal{O}(\alpha^2)$, and
$\mathcal{O}(\alpha^3)$ via photoproduction.}
\label{fig:LHCm}
\end{center}
\end{figure}

\begin{figure}[ht]
\begin{center}
\includegraphics[bb=170 460 425 705,width=0.45\textwidth]%
{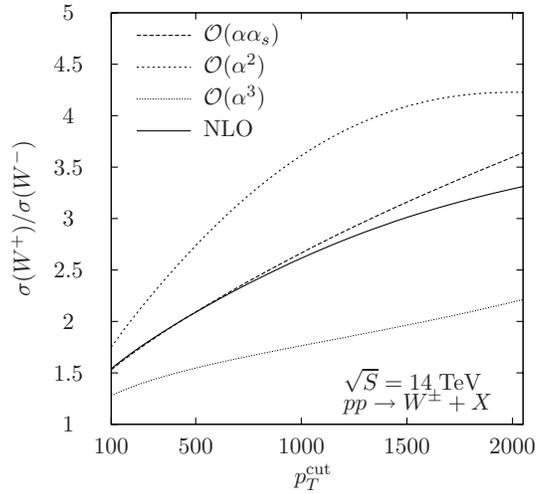}
\caption{Ratios of the respective results for $W^+$ and $W^-$ bosons shown in
Figs.~\ref{fig:LHCp}(a) and \ref{fig:LHCm}(a).}
\label{fig:ptom}
\end{center}
\end{figure}

Let us now discuss the numerical results and their phenomenological
implications in detail.
Specifically, Figs.~\ref{fig:Tev}, \ref{fig:K}(a), and \ref{fig:xi}(a) refer
to the Tevatron, while Figs~\ref{fig:K}(b), \ref{fig:xi}(b), \ref{fig:LHCp},
\ref{fig:LHCm}, and \ref{fig:ptom} refer to the LHC.
In Fig.~\ref{fig:Tev}(a) the NLO result for the total cross section as a
function of $p_T^\mathrm{cut}$ is compared with the LO contributions of
$\mathcal{O}(\alpha\alpha_s)$ and $\mathcal{O}(\alpha^2)$ as well as with the
photoproduction contribution of $\mathcal{O}(\alpha^3)$.
The $\mathcal{O}(\alpha\alpha_s)$ and $\mathcal{O}(\alpha^2)$ results exhibit
very similar line shapes, but the normalisation of the latter is suppressed by
a factor of about 500.
This may be qualitatively understood from the partonic cross section formulae
in Eq.~(\ref{eq:loxs}) and by noticing that the $\mathcal{O}(\alpha\alpha_s)$
contributions from the Compton-like processes~(\ref{eq:ugd}) and
(\ref{eq:dgu}), which have no counterparts in $\mathcal{O}(\alpha^2)$, are
significantly enhanced by the gluon PDF.
As a consequence, the LO result is almost entirely exhausted by the
$\mathcal{O}(\alpha\alpha_s)$ contribution.

The inclusion of the NLO correction leads to a moderate reduction in cross
section, which increases in magnitude with $p_T^\mathrm{cut}$, reaching about
$-4\%$ for $p_T^\mathrm{cut}=200$~GeV, as may be seen from
Fig.~\ref{fig:K}(a), where the $K$ factor is depicted.

In Fig.~\ref{fig:Tev}(a), also the photoproduction contribution is shown.
As explained above, we have to distinguish between elastic and inelastic
scattering off the proton on the one hand, and between direct and resolved
photons on the other hand, so that, altogether, we have four different
contributions, which all formally contribute at $\mathcal{O}(\alpha^3)$.
The resolved-photon contributions turn out to be small against the
direct-photon ones and are, therefore, not included in Fig.~\ref{fig:K}(a).
As for the combined direct-photoproduction contribution, we observe from
Fig.~\ref{fig:Tev}(a), that, except for small values of $p_T^\mathrm{cut}$, it
overshoots the $\mathcal{O}(\alpha^2)$ contribution, although it is formally
suppressed by one power of $\alpha$!
Detailed inspection reveals that this unexpected enhancement can be traced to
the direct-photoproduction diagram involving the triple-gauge-boson coupling
and the space-like $W$-boson exchange, which significantly contributes at
large values of $\sqrt{s}$.
In fact, for a fixed value of $p_T^\mathrm{cut}$, the total cross sections of
processes~(\ref{eq:upd}) and (\ref{eq:dpu}) have an asymptotic large-$s$
behaviour proportional to $1/(m_T^\mathrm{cut})^2$, while those of
processes~(\ref{eq:udg})--(\ref{eq:udp}) behave as $\ln s/s$.
Consequently, photoproduction appreciably contributes to the $K$ factor, as is
apparent from Fig.~\ref{fig:K}(a), which also shows the photoproduction to LO
ratios for elastic and inelastic scattering.
The freedom in the choice of the inelastic photon content of the proton is
likely to be the largest source of theoretical uncertainty in the
photoproduction cross section.
In order to get an idea of this uncertainty, we display in Fig.~\ref{fig:K}(a)
also the inelastic-photoproduction to LO ratio evaluated with the \verb/GSV/
photon spectrum for inelastic scattering.
The result is roughly a factor of two smaller than our default prediction
based on the \verb/MRSTQED04/ spectrum.

In Fig.~\ref{fig:xi}(a), we examine the theoretical uncertainties in the
$\mathcal{O}(\alpha\alpha_s)$, $\mathcal{O}(\alpha^2)$, NLO, and
photoproduction results due to the freedom in setting the renormalisation and
factorisation scales by exhibiting their $\xi$ dependencies relative to their
default values at $\xi=1$.
The $\xi$ dependencies of the $\mathcal{O}(\alpha^2)$ and
direct-photoproduction results stem solely from the factorisation scale $M$
and are rather feeble, while those of the $\mathcal{O}(\alpha\alpha_s)$ and
resolved-photoproduction results are also linked to the renormalisation scale
$\mu$ of $\alpha_s(\mu)$ and are more pronounced, but still not dramatic.
The scale variation of the LO result amounts to less than $\pm15\%$ for
$1/2<\xi<2$.
It is only slightly reduced by the inclusion of the NLO correction.
This is expected because the NLO result is still linear in $\alpha_s(\mu)$, so
that the $\mu$ dependence of $\alpha_s(\mu)$ is not compensated yet.

In Fig.~\ref{fig:Tev}(b), the analysis of Fig.~\ref{fig:Tev}(a) is repeated
for the $p_T$ distribution.
We observe that the line shapes and relative normalisations of the various
distributions are very similar to those in Figs.~\ref{fig:Tev}(a) and the
same comments apply.

Turning to the LHC, we can essentially repeat the above discussion for the
Tevatron, except that we have to take into account the difference between
$W^+$ and $W^-$ boson production.
Thus, Fig.~\ref{fig:Tev} has two LHC counterparts, Figs.~\ref{fig:LHCp} and
\ref{fig:LHCm}, for the $W^+$ and $W^-$ bosons, respectively.
To illustrate this difference more explicitly, we show in Fig.~\ref{fig:ptom}
the $W^+$ to $W^-$ ratios of the respective results from Figs.~\ref{fig:LHCp}
and \ref{fig:LHCm}.
For simplicity, Figs.~\ref{fig:K}(b) and \ref{fig:xi}(b), the LHC counterparts
of Figs.~\ref{fig:K}(a) and \ref{fig:xi}(a), refer to the averages of the
results for $W^+$ and $W^-$ bosons.
In the following, we only focus on those features which are specific for the
LHC.
From Figs.~\ref{fig:LHCp} and \ref{fig:LHCm}, we observe that the gaps between
the $\mathcal{O}(\alpha\alpha_s)$ and $\mathcal{O}(\alpha^2)$ results are
increased by about a factor of two, to reach three orders of magnitude.
This is mainly because the Compton-like processes~(\ref{eq:ugd}) and
(\ref{eq:dgu}) benefit from the extended dominance of the gluon PDF at small
values of $x$.
Furthermore, the photoproduction contributions now significantly exceed the
$\mathcal{O}(\alpha^2)$ ones throughout the entire $p_T^\mathrm{cut}$ and
$p_T$ ranges.
From Fig.~\ref{fig:ptom}, we see that the $W^+$ to $W^-$ ratios take values in
excess of unity, as expected, and strongly increase with increasing values of
$p_T^\mathrm{cut}$.
Comparing Figs.~\ref{fig:K}(a) and (b), we find that the $K$ factors are
significantly amplified as one passes from the Tevatron to the LHC.
This is due to the fact that the Sudakov logarithms, which originate
from triangle and box diagrams, become quite sizeable at the large values of
$\sqrt{s}$ and $p_T$ that can be reached at the LHC.
This issue was already dwelled on in Ref.~\cite{Kuhn:2007qc}, to which we
refer the interested reader.
Finally, comparing Fig.~\ref{fig:xi}(a) and (b), we conclude that the $\xi$
dependence is generally somewhat smaller at the LHC.

\section{Conclusions}

We studied the effect of electroweak radiative corrections at first order on
the cross section of the inclusive hadroproduction of single $W$ bosons with
finite values of $p_T$, putting special emphasis on the notion of
infrared-save observables with a democratic treatment of hadron jets initiated
by (anti)quarks and gluons.
This is indispensable because, as a matter of principle, a collinear
gluon-photon system cannot be distinguished from a single gluon with the same
momentum, so that a minimum-transverse-momentum cut on the gluon is an 
inadequate tool to prevent a soft-gluon singularity.
This led us to include the $\mathcal{O}(\alpha_s)$ correction to $W+\gamma$
production along with the $\mathcal{O}(\alpha)$ correction to $W+j$
production, both contributing at absolute order
$\mathcal{O}(\alpha^2\alpha_s)$.
We also considered the contribution from events where one of the colliding
hadrons interacts via a real photon, which is of absolute order
$\mathcal{O}(\alpha^3)$.
The hadron can then either stay intact (elastic scattering) or be destroyed
(inelastic scattering), and the photon can participate in the hard scattering
directly (direct photoproduction) or via its quark and gluon content (resolved
photoproduction), so that four combinations are possible.

We extracted the UV singularities using dimensional regularisation and removed
them by renormalisation in the on-shell scheme.
We regularised the soft and collinear IR singularities by means of
infinitesimal photon, gluon, and quark masses, $\lambda$, $m_u$, and $m_d$,
respectively.
We used the phase-space slicing method, with cuts $\delta_s$,
$\Delta\vartheta$, and $\Delta\psi$ on the scaled photon and gluon energies
and on the separation angles in the initial and final states, respectively, to
isolate the soft and collinear singularities within the corrections from real
particle radiation.
We achieved the cancellation of $\lambda$, $m_u$, and $m_d$ analytically and
ensured that the numerical results are insensitive to variations of $\delta_s$,
$\Delta\vartheta$, and $\Delta\psi$ within wide ranges about their selected
values.

We presented theoretical predictions for the total cross sections with a
minimum-$p_T$ cut and for the $p_T$ distributions to be measured in
$p\overline{p}$ collisions with $\sqrt{S}=1.96$~TeV at run~II at the Tevatron
and in $pp$ collisions with $\sqrt{S}=14$~TeV at the LHC, and estimated the
theoretical uncertainties from the scale setting ambiguities.
We found that considerably less than 1\% of all $W+X$ events contain a prompt
photon.
The corrections considered turned out to be negative and to increase in
magnitude with the value of $p_T$.
While the reduction is moderate at the Tevatron, reaching about $-4\%$ at
$p_T=200$~GeV, it can be quite sizeable at the LHC, of order $-30\%$ at
$p_T=2$~TeV, which is due to the well-known enhancement by Sudakov logarithms.
It is an interesting new finding that the photoproduction contribution is
considerably larger than expected from the formal order of couplings.
In fact, it compensates an appreciable part of the reduction due to the
$\mathcal{O}(\alpha^2\alpha_s)$ correction.

\section*{Acknowledgement}

We are grateful to Stefan Dittmaier for helpful theoretical and practical
advice, to Gustav Kramer for useful advice regarding phase space slicing, and
to Thomas Hahn, Max Huber, and Frank Fugel for beneficial discussions.
The work of B.A.K. was supported in part by the German Federal Ministry
for Education and Research BMBF through Grant No.\ 05~HT6GUA.

\begin{figure}[ht]
\begin{center}
\input{diagudWg.tex}
\input{Wgammaproddiag.tex}
\end{center}
\caption{Tree-level diagrams of (a) process~(\ref{eq:udg}) and (b)
process~(\ref{eq:udp}).
The tree-level diagrams of processes~(\ref{eq:ugd}), (\ref{eq:dgu}),
(\ref{eq:upd}), and (\ref{eq:dpu}) emerge through crossing.}
\label{DiagBornudWg}
\end{figure}

\end{document}